 \documentclass[12pt]{article}

\usepackage{amsmath}
\usepackage{amssymb}

\begin{document}

\newcommand{\beqn}{\begin{eqnarray}}
\newcommand{\eeqn}{\end{eqnarray}}
\newcommand{\be}{\begin{equation}}
\newcommand{\ee}{\end{equation}}
\newcommand{\ba}{\begin{array}}
\newcommand{\ea}{\end{array}}
\newcommand{\pa}{\partial}
\newcommand{\re}{\ref}
\newcommand{\ci}{\cite}
\newcommand{\la}{\label}
\newcommand{\bfr}{\begin{flushright}}
\newcommand{\efr}{\end{flushright}}
\newcommand{\bfl}{\begin{flushleft}}
\newcommand{\efl}{\end{flushleft}}
\newcommand{\fr}{\frac}
\newcommand{\ov}{\overline}
\newcommand{\ve}{\varepsilon}
\newcommand{\de}{\delta}
\newcommand{\al}{\alpha}
\newcommand{\ga}{\gamma}\newcommand{\Ga}{\Gamma}
\newcommand{\si}{\sigma}
\newcommand{\ds}{\displaystyle}
\newcommand{\pr}{\prime}
\newcommand{\La}{\Lambda}
\newcommand{\Lr}{\Longrightarrow}
\newcommand{\De}{\Delta}
\newcommand{\Si}{\Sigma}
\newcommand{\ti}{\tilde}
\newcommand{\Om}{\Omega}
\newcommand{\om}{\omega}
\newcommand{\na}{\nabla}
\newcommand{\lam}{\lambda}
\newcommand{\Lam}{\Lambda}

\newcommand{\T}{\mathbb{T}}
\newcommand{\R}{\mathbb{R}}
\newcommand{\Z}{\mathbb{Z}}
\newcommand{\N}{\mathbb{N}}
\newcommand{\C}{\mathbb{C}}
\newcommand{\br}{|\kern-.25em|\kern-.25em|}
\renewcommand{\theequation}{\thesection.\arabic{equation}}

\def\Re {{\rm Re\, }}                               
\def\Im {{\rm Tm\,}}                                
\newcommand{\const}{\mathop{\rm const}\nolimits}
\newcommand{\tr}{\mathop{\rm tr}\nolimits}
\newcommand{\supp}{\mathop{\rm supp}\nolimits}
\newcommand{\diam}{\mathop{\rm diam}\nolimits}
\newtheorem{theorem}{Theorem}[section]
\renewcommand{\thetheorem}{\arabic{section}.\arabic{theorem}}
\newtheorem{definition}[theorem]{Definition}
\newtheorem{deflem}[theorem]{Definition and Lemma}
\newtheorem{lemma}[theorem]{Lemma}
\newtheorem{example}[theorem]{Example}
\newtheorem{remark}[theorem]{Remark}
\newtheorem{remarks}[theorem]{Remarks}
\newtheorem{cor}[theorem]{Corollary}
\newtheorem{pro}[theorem]{Proposition}

\newcommand{\bo}{{\hfill\loota}}
\newcommand{\loota}{\hbox{\enspace{\vrule height 7pt depth 0pt width 7pt}}}

\begin{titlepage}
\begin{center}
{\Large\bf Local Stationarity for Lattice Dynamics in the\medskip\\
 Harmonic Approximation}\\
\vspace{2cm}
{\large T.V.~Dudnikova
\footnote{
Supported partly by
research grants of DFG (436 RUS 113/615/0-1)
and RFBR (03-01-00189)
}}\\
{\it  M.V.Keldysh Institute\\
of Applied Mathematics RAS\\
 Moscow 125047, Russia}\\
e-mail:~dudnik@elsite.ru, dudnik@ma.tum.de
 \bigskip\\
 {\large H.~Spohn}\\
{\it Zentrum Mathematik\\
Technische Universit\"at M\"unchen\\
 D-85747 Garching, Germany}\\
 e-mail:~spohn@ma.tum.de
 \end{center}
 \vspace{1cm}
 \begin{abstract}
We consider the lattice dynamics in the harmonic approximation for
a simple hypercubic lattice with arbitrary unit cell. The initial
data are random according to a probability measure which enforces
slow spatial variation on the linear scale $\varepsilon^{-1}$. We
establish two time regimes. For times of order
$\varepsilon^{-\gamma}$, $0<\gamma<1$, locally the measure
converges to a Gaussian measure which is space-time stationary
with a covariance inherited from the initial (in general,
non-Gaussian) measure. For times of order $\varepsilon^{-1}$ this
local space covariance changes in time and is governed by a
semiclassical transport equation.\bigskip
 \end{abstract}
\end{titlepage}

 \section{Introduction}

For systems consisting of many interacting ``particles" as a rule the slow
degrees of freedom are linked to local conservation laws.
For example for a classical fluid, mass, momentum, and energy are
locally conserved and as a consequence mass, momentum, and energy
density are the slow degrees of freedom. Thus if the system starts
with some general initial conditions, one expects the fast degrees
of freedom to die out rapidly. Then, in a spatial region which on
one side contains many particles and on the other side is still
small compared to the total extent of the system, thus
\textit{locally}, the statistical distribution on phase space is
stationary under the dynamics within a good approximation. Since
the system has not yet reached global stationarity, there is still
a slow motion of the parameters characterizing the states of local
stationarity. In our example of a classical fluid, local
stationarity coincides with local thermal equilibrium and the
local equilibrium parameters, density, momentum, and internal
energy, evolve according to the Euler equations of fluid dynamics.
For other systems with many particles, in general, it is a
difficult task to identify the relevant probability measures
stationary in time (and usually also in space).

Such a picture for the dynamics of systems with many particles has
theoretical and mathematical support. We refer to \cite{Sp91}. If
the dynamics is of Hamiltonian form, the list of worked out
examples is rather short. One item on the list is lattice dynamics
in the harmonic approximation, which has been investigated in
great detail by R.L. Dobrushin and collaborators \cite{DPST}. We
reconsider this model for two reasons.\smallskip

(i) The first one is on a conceptual level. In phonon physics it
is standard practice to use the Wigner function $W(t,r,\theta)$ as
density of phonons with wave number $\theta$ at location $r$ and
at specified time $t$. $W$ evolves according to the semiclassical
transport equation
 \be\la{I.1}
 \frac{\partial}{\partial t}W(t,r,\theta)= -\nabla\omega(\theta)
 \nabla_r W(t,r,\theta)\,,
\ee $\omega(\theta)$ being the dispersion relation of the harmonic
crystal. As we will establish,
$W(t,r,\theta)\delta(\theta-\theta')$ at fixed $r,t$ encodes the
covariance of a Gaussian measure on phase space which is invariant
under the lattice dynamics. Thus (\ref{I.1}) can be understood as
the equation governing the motion of the parameters which
characterize the locally stationary measures. We believe that in
this way the results of Dobrushin \textit{et al.} become more
transparent and, in addition, the link to the physics of phonons
is provided, see \cite{Sp05} for a more detailed
discussion.\smallskip

(ii) The second reason is technically. In the recent years there
has been considerable progress in understanding the long time
limit of the harmonic crystal in infinite volume \cite{DKS1}.
More precisely one starts with a probability measure $\mu_0$ which
is translation invariant and has some mixing properties. If
$\mu_t$ denotes the time-evolved measure at time $t$, then the
limit
 \be\la{I.2}
 \lim_{t\to\infty} \mu_t=\mu_\infty\,,
\ee is established, where $\mu_\infty$ is a suitable Gaussian
measure with mean zero. It turns out that the techniques for
proving (\ref{I.2}) transcribe to the locally stationary
situation. Thereby the conditions in the work of Dobrushin
\textit{et al.} are considerably streamlined and the proof is
simplified. We also generalize from one to an arbitrary space
dimension and from one particle per unit cell to an arbitrary
number.

In a recent paper \cite{M}, A.~Mielke studies the same model and
also obtains the semiclassical transport equation (\ref{I.1}) for
the Wigner function. However, Mielke imposes deterministic initial
data of slow variation, while we impose random initial data with
rather strong mixing properties. Therefore the results are
disjoint and so are the techniques for proving them. It is of
interest to understand whether a ``supertheorem" encompassing both
cases has a chance to be valid.

\setcounter{equation}{0}
 \section{Lattice dynamics in the harmonic approximation}
\subsection{The model}

We consider a Bravais lattice with a unit cell which contains a
finite number of atoms. For notational simplicity the Bravais
lattice is assumed to be simple hypercubic. Let $x\in\Z^d$ and let
$u(x)$ be the field of displacements in cell $x$ from the
equilibrium position. If $u$ is small, we may expand the forces to
linear order, which then yields the linear$n\mathrm{-component}$
discrete wave equation
 \be\la{1.1'}
\ddot u(x,t)  =  -\sum_{y\in\Z^d}
  V(x-y) u(y,t),\,\,\,\,\,
u(x)|_{t=0} = u_{0}(x),~~
\dot u(x)|_{t=0} = v_{0}(x), \,\,\,\,x\in\Z^d. \ee
Here
 $u(x,t)=(u_1(x,t),\dots,u_n(x,t)),
 u_0=(u_{01}(x),\dots,u_{0n}(x))\in\R^n$ and
correspondingly for $v_0(x)$. Physically $n=d\times\!$(number of
atoms in the unit cell). Here we take $n$ to be an arbitrary
positive integer. $V(x)$ is an $n\times n$ matrix. The dynamics
(\re{1.1'}) is invariant under lattice translations.

Let us denote by $Y(t)=(Y^0(t),Y^1(t))= (u(\cdot,t),\dot
u(\cdot,t))$, $Y_0=(Y^0_0,Y^1_0)= (u_0(\cdot),v_0(\cdot))$. Then
(\ref{1.1'}) takes the form of an evolution equation
\be\la{CP}
\dot Y(t)={\cal A}Y(t),\,\,\,t\in\R,\,\,\,\,Y(0)=Y_0. \ee
Formally, this is a linear Hamiltonian system, since
 \be\la{A}
{\cal A}Y=J\left(
 \begin{array}{cc}
{\cal V}& 0\\
0 & 1 \end{array}\right)Y
=J\na H(Y),\,\,\,\,\,\,\,\,\,
J= \left(\begin{array}{cc}
0 & 1\\
- 1 & 0\end{array}\right), \ee
with the Hamiltonian
functional
 \be\la{H} H(Y)= \frac{1}{2} \langle v,v\rangle
+\frac{1}{2} \langle  {\cal V}u, u\rangle, \quad Y=(u,v), \ee
where ${\cal V}$ is the convolution
operator with the matrix kernel $V$, the kinetic energy is given by
$\ds\frac{1}{2} \langle v,v\rangle=
\frac{1}{2} \sum_{x\in\Z^d}|v(x)|^2$, and the potential
energy by $\ds\frac{1}{2}\langle{\cal V}  u,u\rangle
=\frac{1}{2}\sum_{x,y\in\Z^d} u(x)\cdot V(x-y)u(y)$. Here
``$\cdot$'' stands for the  scalar product in the
Euclidean space $\R^n\!$, resp. in $\R^d\!$.

 We assume that the initial datum $Y_0$
belongs to the phase space ${\cal H}_\al$ for some
 $\al\in\R$.
 \begin{definition} \la{d1.1}
  $ {\cal H}_\al$  is the  Hilbert space
of pairs $Y=(u,v)$  of  $\R^n\!$-{\textit{valued}} functions
  on $\mathbb{Z}^d$  equipped with the norm
 \beqn                              \la{1.5}
 \Vert Y\Vert^2_{\al}
 =   \sum_{x\in\Z^d}\Big(
\vert u(x)\vert^2
 +  \vert v(x)\vert^2\Big)(1+|x|^2)^{\al} <\infty\,.
 \eeqn
 $ {\cal H}_\al$ is equipped with the Borel $\sigma$-algebra ${\cal B}( {\cal H}_\al)$.
 \end{definition}

We impose the following conditions on  the matrix $V\!$.
\medskip\\
{\bf E1} There exist constants $C,\alpha>0$ such that $\|V(z)\|\le
C e^{-\alpha|z|}$ for $z\in \Z^d\!$, $\|V(z)\|$ denoting the
matrix norm.
\medskip

Let $\hat V(\theta)$ be the Fourier transform of $V(x)$, with the
convention
 \be
 \hat V(\theta)=
\sum\limits_{z\in\Z^d}V(z)e^{iz\cdot\theta}\,,\;\theta \in \T^d\,,
 \ee
$\T^d$ the $d$-torus $\R^d/(2\pi \Z)^d$.
\medskip\\
{\bf E2} $ V$ is even, in the sense that $V(-z)=V(z)^\ast\in \R$,
for $z\in \Z^d$, where $V^\ast$ denotes the adjoint of the matrix
$V$ as acting on $\C^n$.
\medskip\\
Both conditions imply that $\hat V(\theta)$ is
 a real-analytic
 Hermitian matrix-valued function in $\theta\in \T^d\!$.
\medskip\\
{\bf E3} The matrix $\hat V(\theta)$ is  non-negative definite for
every $\theta \in \T^d.$
\medskip

Let us define the Hermitian  non-negative definite matrix
 \be\la{Omega}
 \Omega(\theta)=\big(\hat V(\theta )\big)^{1/2}\ge 0\,.
 \ee
$\Om(\theta)$  has the eigenvalues $0\leq\om_1(\theta)<
\om_2(\theta) \ldots <\om_s(\theta)$, $s\leq n$ and the
corresponding spectral projections $\Pi_\sigma(\theta)$ with
multiplicity $r_\sigma=\mathrm{tr}\Pi_\sigma(\theta)$. $\theta
\mapsto\omega_\sigma(\theta)$ is the $\sigma\!$-th band function.
There are special points in $\mathbb{T}^d$,
where the bands cross, which means that $s$ and $r_\sigma$ jump to
some other value. Away from such crossing points $s$ and
$r_\sigma$ are independent of $\theta$. More precisely one has the
following lemma.

\begin{lemma}\la{lc*} (see \ci[Lemma 2.2]{DKS1}).
Let the conditions {\bf E1}, {\bf E2}
hold. Then there exists a closed subset ${\cal C}_*\subset \T^d$
such that\\
i) the Lebesgue measure of ${\cal C}_*$ is zero.\\
ii) For every point $\Theta\in \T^d\setminus{\cal C}_*$ there
exists a neighborhood ${\cal O}(\Theta)$ such that each band
function $\om_\sigma(\theta)$
can be chosen as real-analytic function in
${\cal O}(\Theta)$.\\
iii) The eigenvalue $\om_\sigma(\theta)$ has constant multiplicity
in
$\T^d\setminus{\cal C}_*$.\\
iv) For $\theta\in \T^d\setminus{\cal C}_*$, the spectral
decomposition
 \be\la{spd'}
\Om(\theta)=\sum_{\sigma=1}^s \om_\sigma
(\theta)\Pi_\sigma(\theta)
 \ee
holds, where $\Pi_\sigma(\theta)$ is an orthogonal projection in
$\R^n\!$. $\Pi_\sigma$ is a real-analytic function on
$\mathbb{T}^d\setminus{\cal C}_*$.
\end{lemma}

For $\theta\in \mathbb{T}^d\setminus{\cal C}_*$ we denote by
Hess$(\omega_\sigma)$ the matrix of second partial derivatives. Our next
condition is the following.
\smallskip\\
{\bf E4} Let
$D_\sigma(\theta)=\det\big(\mathrm{Hess}(\omega_\sigma(\theta))\big)$.
Then
$D_\sigma$ does not vanish identically on
$\T^d\setminus{\cal C}_*$, $\sigma=1,\ldots,s$.
\medskip\\
Let
 \be\la{c0ck}
{\cal C}_0=\{\theta\in \T^d:\det \hat
 V(\theta)=0\}\,\, \mbox{and }\,
{\cal C}_\sigma=\{\theta\in
\T^d\setminus {\cal
C}_*:\,\det(\mathrm{Hess}(\omega_\sigma))=0\},\,\,
\sigma=1,\dots,s.
 \ee
The following lemma has been proved in \ci[Appendix]{DKS1}.
\begin{lemma}\la{lc}
Let the conditions {\bf E1} - {\bf E4} hold. Then the Lebesgue
measure of ${\cal C}_k$ vanishes, $k=0,1,...,s.$
\end{lemma}
Our final conditions on $V$ are the following:
\medskip\\
{\bf E5} For each $\si\ne \si'$, $\om_\si \pm \om_{\si'}$ does not take a
constant value on $\theta\in \T^d\setminus {\cal C}_*$.\medskip\\
This condition holds trivially  in  case $n=1$.
\medskip\\
{\bf E6} $\Vert \hat V^{-1}(\theta)\Vert\in L^1(\T^d)$.\medskip\\
If ${\cal C}_0=\emptyset$, then $\|\hat{V}^{-1}(\theta)\|$ is
bounded and {\bf E6} holds trivially.

\begin{remark}
{\rm The conditions {\bf E1} - {\bf E6} are fairly general. In
particular they can be checked for the case of nearest neighbor
coupling only, for which
 \beqn\la{dKG}
  \langle {\cal V}u,u \rangle=
\sum\limits_{k=1}^n\sum\limits_{x\in\Z^d}
\Big(\sum\limits_{i=1}^{d}\gamma_k |u_k(x+e_i)-u_k(x)|^2+
m_k^2|u_k(x)|^2\Big), \; \gamma_k>0, \,\,\,m_k\geq 0\,,
 \eeqn
  where
$e_i=(\de_{i1},\dots, \de_{id})$. Then the eigenvalues of  $\hat
V(\theta)$ are
 \be\la{omega}
 \tilde{\omega}_k(\theta)=
\sqrt{\, 2 \gamma_1(1-\cos\theta_1)+...+2 \gamma_d
(1-\cos\theta_d)+m_k^2}\,.
 \ee
These eigenvalues still have to be labelled according to magnitude
and degeneracy as in Lemma \ref{lc*}.
Clearly {\bf {E1}} - {\bf {E5}} hold.
In case all $m_k>0$ the set ${\cal C}_0$
 is empty and
condition {\bf E6} holds automatically. Otherwise, if $m_k=0$ for
some $k$,  ${\cal C}_0=\{0\}$. Then
{\bf E6} is equivalent to the condition $\om_k^{-2}(\theta)\in
L^1(\T^d)$, which holds if $d\ge 3$. Therefore, the conditions {\bf
E1} - {\bf E6} hold  for (\ref{dKG}) provided either i) $d\ge 3$,
or ii) $d=1,2$ and all $m_k >0$. }
\end{remark}

\begin{pro} (see \ci[Proposition 2.5]{DKS1}). \la{p1.1}
Let  {\bf E1} and {\bf E2} hold
and choose some $\al\in\R$. Then \\
i) for any  $Y_0 \in {\cal H}_\al$
 there exists  a unique solution
$Y(t)\in C(\R, {\cal H}_\al)$
 to the Cauchy problem (\re{CP}).\\
ii) The operator $U(t):Y_0\mapsto Y(t)$ is continuous
in ${\cal H}_\al$,
$
\Vert U(t) Y_0\Vert_{\al}\le C(t)\Vert Y_0\Vert_{\al}.
$
\end{pro}

\subsection{Random initial data}
We assume that $Y_0$ is a random function with distribution
$\mu_0$.
\begin{definition}  \la{dmut}
$\mu_t$ is a Borel probability measure in ${\cal H}_\al$ which
gives the distribution of $Y(t)$,
 \beqn\nonumber \mu_t(B) =
\mu_0(U(-t)B),\,\,\,\,  B\in {\cal B}({\cal H}_\al), \,\,\,
t\in \R\,.
 \eeqn
Expectation with respect to $\mu_t$ is denoted by
$\mathbb{E}_t\!$.
\end{definition}
We set ${\cal D}=D\oplus D$ with $D= C_0(\Z^d)\otimes\R^n$, where
$C_0(\Z^d)$ denotes a space of real sequences with finite support,
and $\langle Y,\Psi\rangle =\langle Y^0,\Psi^0\rangle +\langle
Y^1,\Psi^1\rangle$ for  $Y=(Y^0,Y^1)\in {\cal H}_\al$ and $
\Psi=(\Psi^0,\Psi^1)\in  {\cal D}$. For a probability  measure
$\mu$ on  ${\cal H}_\al$ we denote by $\hat\mu$ the characteristic
functional (Fourier transform),
$$
\hat \mu(\Psi ) = \int\exp(i\langle Y,\Psi \rangle )\,\mu(dY),
\,\,\, \Psi\in  {\cal D}.
$$
A measure $\mu$ is called Gaussian of zero mean, if its
characteristic functional has the form \beqn\nonumber
\ds\hat { \mu} (\Psi ) =  \ds \exp\big[-\fr12
 {\cal Q}(\Psi , \Psi )\big]\,,\,\,\,\Psi \in {\cal D},
\eeqn where ${\cal Q}$ is a real non-negative quadratic form on
${\cal D}$. A measure $\mu$ is called translation invariant if $
\mu(T_h B)= \mu(B),\,\,\, B\in{\cal B}({\cal H}_\alpha), \,\,\,\,
h\in\Z^d,$ where $T_h Y(x)= Y(x-h)$, $x\in\Z^d$.
\medskip

Let $O(r)$ denote the set of all pairs of  subsets ${\cal
A},\>{\cal B}\subset \Z^d$ at a distance dist$({\cal A},\,{\cal
B})\geq r$ and let $\sigma ({\cal A})$ be the $\sigma $-algebra in
${\cal H}_\al$ generated by $Y(x)$ with $ x\in{\cal A}$. Define
the Ibragimov-Linnik mixing coefficient of a probability  measure
$\mu$ on ${\cal H}_\al$ by (cf. \ci[Definition 17.2.2]{IL})
 \beqn
\la{ilc} \varphi(r)= \sup_{({\cal A},{\cal B})\in O(r)}
\sup_{\scriptsize{
\ba{cc} A\in\si({\cal A}),B\in\si({\cal B})\\ \mu(B)>0\ea}}
\fr{|\mu(A\cap B) - \mu(A)\mu(B)|}{ \mu(B)}.
 \eeqn
\begin{definition}
 A measure $\mu$ satisfies the strong uniform
Ibragimov-Linnik mixing condition if
$
\varphi(r)\to 0$ as $r\to\infty$.
\end{definition}

\setcounter{equation}{0}
 \section{Main results}
\subsection{Spatially homogeneous initial measure}\la{sec.1}

In this subsection we assume that the initial measure, $\mu_0$, is
spatially translation invariant with the following properties.
\medskip\\
{\bf S1} $Y_0(x)$ has zero expectation value,
 \beqn\nonumber
\mathbb{E}_0 \big(Y_0(x)\big) = 0\,,\quad x\in\Z^d\,.
 \eeqn
   {\bf S2} $\mu_0$ has translation invariant
correlation matrices, i.e.,  for $i,j=0,1$,
 \beqn \la{1.9'}
Q^{ij}_0(x,x')= \mathbb{E}_0\big(Y^i_0(x)\otimes {Y^j_0(x')}
\big)= q^{ij}_0(x-x'),\,\,\,x,x'\in\Z^d.
 \eeqn
 Here for $a,b,c \in \mathbb{C}^n$ we denote by $a\otimes b$ the
 linear operator $(a\otimes b)c=a\sum^n_{j=1}b_j c_j$.
  \medskip\\
   {\bf S3} $\mu_0$  has a
finite variance and finite mean energy density,
 \beqn\nonumber
e_0=\mathbb{E}_0 \big(\vert Y_0^0(x)\vert^2
 + \vert Y_0^1(x)\vert^2\big)={\rm tr}\,q_0^{00}(0)+
{\rm tr}\,q_0^{11}(0) <\infty,\,\,\,x\in\Z^d.
 \eeqn
{\bf S4}
 $\mu_0$ satisfies the strong uniform
Ibragimov-Linnik mixing condition with
 \be\la{1.12}
\int\limits_{0}^{\infty}
 r^{d-1}\varphi^{1/2}(r)\,dr <\infty\,.
 \ee

In \ci{DKS1} we prove the weak convergence of the measures $\mu_t$
to a limit measure $\mu_\infty$ on the Hilbert space ${\cal
H}_\alpha$ with $\al<-d/2$, which means
 \be\la{2.2a}
 \lim_{t\to\infty}\int
f(Y)\,\mu_t(dY)=\int f(Y)\,\mu_\infty(dY)
 \ee
for all bounded continuous functions $f$ on ${\cal H}_{\al}$.
$\mu_{\infty}$ is a Gaussian measure on ${\cal H}_\alpha$.
\begin{theorem} \la{the1}  (see \ci{DKS1}).
{\it  Let $d,n\ge 1$,
$\al<-d/2$, and assume that the conditions {\bf E1} - {\bf E6} and
{\bf S1} - {\bf S4} hold. Then
\smallskip\\
i) the correlation matrices of the measures $\mu_t$
converge to a limit, for $i,j=0,1$,
$$
Q^{ij}_t(x,x')=\int\big( Y^i(x)\otimes Y^j(x')\big)
\,\mu_t(dY)\to Q^{ij}_\infty(x,x'),\,\,\,\,t\to\infty.
$$
ii) The convergence in (\re{2.2a}) holds.\smallskip\\
iii) The limit measure
$ \mu_\infty $ is a Gaussian
measure on ${\cal H}_\al$.\smallskip\\
iv) The  correlation matrix of $ \mu_{\infty}$ is translation
invariant, $Q_\infty(x,x')=q_\infty(x-x')$, and has the Fourier
transform
 \beqn\nonumber
\hat q_\infty(\theta)= \sum_{\si=1}^s
\Pi_\si(\theta)M_0(\theta)\Pi_\si(\theta),
 \,\,\,\,\,\,\theta\in\T^d\setminus{\cal C}_*\,,
 \eeqn
where $\Pi_\si(\theta)$ is the spectral projection from Lemma
\re{lc*} iv) and
\beqn\nonumber
 M_{0}(\theta)
=\frac{1}{2} \left(\hat q_0(\theta) +C(\theta)\hat
q_0(\theta)C(\theta)^*\right) \eeqn with \be\la{C(theta)}
C(\theta)=\left(\ba{cc}
0&\Omega(\theta)^{-1}\\
-\Omega(\theta)&0 \ea\right)\,.
 \ee
v) The measure $\mu_\infty$ is time stationary, i.e.
$[U(t)]^*\mu_\infty=\mu_\infty$, $t\in\R$.}
\end{theorem}

The projection of the initial covariance $\widehat{q}_0(\theta)$
to the limiting covariance $\widehat{q}_\infty(\theta)$ can be
stated more concisely through introducing the complex-valued field
 \beqn\nonumber
 a(x)= \frac{1}{\sqrt{2}}\Big({\cal V}^{1/4}
 u(x)+i{\cal V}^{-1/4}v(x)\Big)\in \mathbb{C}^n\,,
\quad x\in \mathbb{Z}^d\,,
 \eeqn
with complex conjugate field $a(x)^*$ and distributional Fourier transform
$\hat a(\theta)$.

Obviously $\mathbb{E}_t \big(a(x)\big)=0$. The covariance has two
parts. By Theorem \ref{the1} the $aa$-, equivalently the $a^\ast
a^\ast$-, covariance satisfies
 \beqn\nonumber
 \lim_{t\to\infty}\mathbb{E}_t \big(a(x)\otimes a(x')\big)=0\,.
 \eeqn
For the $a^\ast a$-covariance we define
 \beqn\nonumber
 \mathbb{E}_t \big(\hat{a}(\theta)^\ast\otimes \hat{a}(\theta')\big)=
  (2\pi)^d\delta(\theta-\theta')W(t,\theta)\,,
 \eeqn
using the translation invariance of $\mu_t$. Note that
$W(t,\theta)\geq0$. Then
 \beqn\la{2.5}
 \lim_{t\to\infty}W(t,\theta)=\sum^s_{\sigma=1}\Pi_\sigma(\theta)
 W(0,\theta)\Pi_\sigma(\theta)\,.
 \eeqn

\subsection{Initial measure with slow variation}\la{sec.2}

Let $\{\mu^\varepsilon_0,\varepsilon>0\}$ be a family of initial
measures. Roughly, in a linear region of size $\varepsilon^{-1}$,
$\varepsilon\ll 1$, $\mu^\varepsilon_0$ looks like the spatially
homogeneous initial measure from Section \ref{sec.1}. However the
covariance $Q^{ij}_0$ depends on the spatial region under consideration,
 and not only on the difference $x-x'$.

To be more precise let us introduce the complex $2n\times 2n$
matrix-valued function $\hat R$ on $\R^d\times \T^d$, through
\beqn\nonumber
\hat R(r,\theta)= \left(\ba{ll}
\hat R^{00}(r,\theta)&\hat R^{01}(r,\theta)\\
\hat R^{10}(r,\theta)&\hat R^{11}(r,\theta)
\ea\right),\,\,\,\,r\in \R^d,\,\,\,\,\theta\in \T^d, \eeqn with
the following properties.
\medskip\\
{\bf I1} For every fixed $r\in\R^d$ and $i,j=0,1$, the entries of
the matrix-valued function $\hat{R}$ are bounded on $\T^d$ and the
inverse Fourier transform
$$
R^{ij}(r,x)=(2\pi)^{-d}\int\limits_{\T^d} e^{-i\theta\cdot x}\hat
R^{ij}(r,\theta)\,d\theta
$$
satisfies the bound
\be\la{3.2} |R^{ij}(r,x)|\le
C(1+|x|)^{-\gamma},\,\,\,\,x\in\Z^d,
 \ee
where $C$ is some  positive constant, $\gamma >d$.
\medskip\\
{\bf I2} For every fixed $r\in\R^d$, the matrix-valued function
$\hat{R}$ satisfies
 \be\la{3.3}
 \hat R^{00}(r,\theta)\geq 0\,,\quad \hat R^{11}(r,\theta)\geq 0
\ee
$$\hat R^{01}(r,\theta)=\hat R^{10}(r,\theta)^*,
\,\,\,\,\theta\in \T^d.
$$
{\bf I3}  For every fixed $r\in\R^d$ and $\theta\in \T^d$,
the matrix $\hat R(r,\theta)$ is non-negative definite.\medskip\\
{\bf I4}  For every $\theta\in \T^d$, $\hat R^{ij}(\cdot,\theta)$,
$i,j=0,1$, are $C^1$ functions and the function
$$
r\to \sup_{\theta\in \T^d}\max_{i,j=0,1}
\big(\big|\hat R^{ij}(r,\theta)\big|,
\big|\nabla_r \hat R^{ij}(r,\theta)\big|\big)
$$
is bounded uniformly on bounded sets.
\medskip

Let $\mathbb{E}^\ve_0$ stand for expectation w.r.t.~the measure
$\mu_0^\ve$. We assume that
 \be
 \mathbb{E}^\varepsilon_0\big(Y^j(x)\big)=0
 \ee
and define the covariance
 \beqn\nonumber
 Q^{ij}_{\ve}(x,x')= \mathbb{E}^\ve_0\big(Y^i(x)\otimes
Y^j(x')\big),\,\,\,\,x,x'\in \Z^d,\,\,\,\,i,j=0,1\,.
 \eeqn
\begin{definition}
We call a family of measures $\{\mu_0^\ve,\ve>0\}$
 a family of slow variation  for $R$ if
$\{Q^{ij}_{\ve}(x,x'),\ve>0\}$ satisfies the conditions
{\bf V1} - {\bf V4} listed below.
\medskip\\
{\rm {\bf V1} For any $\ve >0$ there exists an  even integer $N_\ve$
such that\medskip\\
i) for all $M\in \R^d$ and $x,x'\in I_{M}$,
\beqn\la{2.4}
\left|Q^{ij}_{\ve}(x,x')- R^{ij}(\ve M,x-x')\right|\le
C\min[(1+|x-x'|)^{-\gamma}, \ve N_\ve],
\eeqn
where $C$, $\gamma$
are the constants from (\ref{3.2}), and $I_{M}$ is the cube
centered at the point $M$ with edge length $N_\ve$,
 \be \la{cube}
I_{M} =\{x=(x_1,\dots,x_d)\in\Z^d:\, |x_j-M_j|\le
N_\ve/2,\,M=(M_1,\dots,M_d)\}.
 \ee
ii) $N_\ve\sim \ve^{-\beta}$ as $\ve\to 0$, with some
$\beta\in(1/2,1)$.\medskip\\
{\bf V2}
 For any $\ve>0$ and all $x,x'\in\Z^d$, $i,j=0,1$,
\beqn\nonumber
|Q^{ij}_{\ve}(x,x')|\le C(1+|x-x'|)^{-\gamma}
\eeqn
with constants $C\,,\,\gamma$ as in (\ref{3.2}).
\medskip\\
{\bf V3} For any $\ve >0$ and any $\Psi_1, \Psi_2\in{\cal D}$ with
dist$(\supp \Psi_1,\supp \Psi_2)\ge\rho>0$ there exist constants
$C>0$ and $\kappa\in(0,1)$
 such that
\beqn\nonumber \left| \mathbb{E}^\ve_0 \big(e^{i\langle
Y,\Psi_1\rangle} e^{i\langle Y,\Psi_2\rangle}\big)
-\mathbb{E}^\ve_0 \big(e^{i\langle Y,\Psi_1\rangle}\big)
\mathbb{E}^\ve_0
\big(e^{i\langle Y,\Psi_2\rangle}\big) \right| \le C(1+\rho)^{-\kappa}.
\eeqn
{\bf V4} For correlation functions of the fourth order
$$
M^{(4)}_\ve(x^1,x^2,x^3,x^4) = \mathbb{E}^\ve_0\big(Y(x^1)\otimes
Y(x^2)\otimes Y(x^3)\otimes Y(x^4)
\big),\,\,\,\,x^1,\dots,x^4\in\Z^d,
$$
we require that
\beqn\nonumber |M^{(4)}_\ve(x^1,x^2,x^3,x^4)|\le C
\sum\limits_{(i_1,i_2,i_3,i_4)\in P\{1,2,3,4\}}
(1+|x^{i_1}-x^{i_2}|)^{-\gamma}(1+|x^{i_3}-x^{i_4}|)^{-\gamma},
\eeqn where $P\{1,2,3,4\}$ is a permutation of the numbers $1,2,3,4$,
and $\gamma>d$.}
\end{definition}
\begin{definition}
i) $\mu^\ve_t$ is a Borel probability measure in ${\cal H}_\al$
which gives the joint distribution of $Y(t)$,
\begin{eqnarray} \nonumber
\mu^\ve_t(B) = \mu_0^\ve(U(-t)B),\,\,\,\,  B\in {\cal
B}({\cal H}_\al), \,\,\,   t\in \R\,.
 \eeqn ii) The correlation
functions of the  measure $\mu^\ve_t$
 are  defined by
\beqn\nonumber
Q_{\ve,t}^{ij}(x,y)=\int Y^i(x)\otimes
 Y^j(y)\mu^\ve_{t}(dY)= \mathbb{E}^{\ve}_0\big(Y^i(x,t)\otimes
 Y^j(y,t)\big),
\,\,\,i,j= 0,1,\,\,\,\,x,y\in\Z^d. \eeqn Here $Y^i(x,t)$ are the
components of the solution $Y(t)=(Y^0(\cdot,t),Y^1(\cdot,t))$.
\end{definition}

\subsection{Covariance in the kinetic scaling limit}

The family $\mu^\varepsilon_0$, $\varepsilon>0$, of initial
measures has slow spatial variation on scale $\varepsilon^{-1}$
and for long times, roughly of order $\varepsilon^{-\gamma}$,
$0<\gamma<1$, in essence Theorem \ref{the1} applies locally, which
implies that locally the projected measure is attained. This
measure is then almost invariant under the time evolution. Thus
one needs a time span of order $\tau/\varepsilon$, $\tau\neq 0$,
to see changes in the projected part of the covariance.

To state a precise result we introduce the scaled $n\times n$ Wigner
matrix through
 \be\la{3.15}
W^\varepsilon(\tau;r,\theta)=
\sum_{y\in \mathbb{Z}^d}e^{i\theta\cdot y}\,
\mathbb{E}^\varepsilon_{\tau/\varepsilon}
\big(a^\ast([\varepsilon^{-1}r+y/2])
\otimes a([\varepsilon^{-1} r-y/2])\big)\,.
 \ee
By our assumptions on $\mu^\varepsilon_0$, the following limit
exists
 \beqn\la{3.12'}
 \lim_{\varepsilon\to 0}W^\varepsilon(0;r,\theta)&=&
 \frac{1}{2}\Big(\Omega^{1/2}\hat R^{00}(r,\theta)\Omega^{1/2}+
 \Omega^{-1/2}\hat R^{11}(r,\theta)\Omega^{-1/2}\nonumber\\
 &&+i\Omega^{1/2}\hat R^{01}(r,\theta)\Omega^{-1/2}-i
 \Omega^{-1/2}\hat R^{10}(r,\theta)\Omega^{1/2}\Big)\nonumber\\
 &=& W(0;r,\theta)\,.\nonumber
 \eeqn
We also define the projected initial Wigner matrix, compare with
(\ref{2.5}),
 \be\la{3.16}
 W^\mathrm{p}(r,\theta)=\sum^s_{\sigma=1}
 \Pi_\sigma(\theta)W(0;r,\theta)\Pi_\sigma(\theta)
 \ee
and its time evolution
 \be\la{3.16a}
 W^\mathrm{p}(\tau;r,\theta)=\sum^s_{\sigma=1}
 \Pi_\sigma(\theta)W(0;r-\tau\nabla\omega_\sigma(\theta),\theta)
 \Pi_\sigma(\theta)\,.
 \ee
\begin{theorem} \la{the2}
Let the conditions {\bf{V1}} - {\bf{V2}} and {\bf{E1}} - {\bf{E6}}
hold. Then for any $r\in\R^d$ and
$\tau\neq 0$ the following limit exists
in the sense of distributions,
 \be\la{3.17}
 \lim_{\varepsilon\to 0} W^\varepsilon(\tau;r,\theta)=
 W^\mathrm{p}(\tau;r,\theta)\,.
 \ee
In addition, for the remaining part of the covariance,
 \be
 \lim_{\varepsilon\to 0}\sum_{y\in\mathbb{Z}^d}e^{i\theta\cdot
 y}\,\mathbb{E}^\varepsilon_{\tau/\varepsilon}
\big(a([\varepsilon^{-1}r+y/2])\otimes
 a([\varepsilon^{-1}r-y/2])\big)=0\,.
 \ee
\end{theorem}

We remark that in the $\sigma$-th band the Wigner function evolves
according to the transport equation
 \be\la{3.18}
 \frac{\partial}{\partial
 t}f_t(r,\theta)+\nabla\omega_\sigma(\theta)\cdot \nabla_r
 f_t(r,\theta)=0\,,
 \ee
where the initial conditions are given by the initial Wigner
matrix projected onto the $\sigma$-th band.

The conditions {\bf{V1}} and {\bf{V2}} on the initial measure are
written in position space. Therefore it is natural to prove the
limiting covariance first in position space, which will be stated
in Theorem \ref{the2'}. From it we deduce the limiting Wigner function of
Theorem \ref{the2}.

\subsection{Local stationarity}

So far we studied only the covariance. A more detailed statistical
information is provided by considering the random field $Y$ at the
kinetic time $\tau/\varepsilon$, $\tau\neq 0$, and close to the
spatial point $[r/\varepsilon]\in \mathbb{Z}^d$. For this purpose
let $T_h$, $h\in \mathbb{Z}^d$, be the group of space
translations. The measure at $r/\varepsilon$ is then defined
through
 \be\la{13.19}
 \mu^\varepsilon_{\tau/\varepsilon,r}=T_{-[r/\varepsilon]}
 \mu^\varepsilon_{\tau/\varepsilon}\,.
 \ee
\begin{theorem} \la{the3}
Let the conditions {\bf{V1}} - {\bf{V4}} and {\bf{E1}} - {\bf{E6}}
hold. Then for $\tau\neq 0$, in the sense of weak convergence on
$\mathcal{H}_\alpha$,
 \be\la{13.20}
 \lim_{\varepsilon\to
 0}\mu^\varepsilon_{\tau/\varepsilon,r}=\mu^\mathrm{G}_{\tau,r}\,.
 \ee
 $\mu^\mathrm{G}_{\tau,r}$ is a Gaussian measure on $\mathcal{H}_\alpha$,
which is invariant under the space translations $T_h$ and time
translation $U(t)$. $\mu^\mathrm{G}_{\tau,r}$ has mean zero and
covariance
 \beqn\nonumber
 q^{ij}_{\tau,r}(x-x')=
 \mathbb{E}^\mathrm{G}_{\tau,r}\big(Y^i(x)\otimes Y^j(x')\big)\,,
 \eeqn
 expectation with respect to  $\mu^\mathrm{G}_{\tau,r}$. The
 covariance is determined through $W^\mathrm{p}(\tau;r,\theta)$ as
  \be\la{13.21}
  \Omega(\theta)\hat{q}^{00}_{\tau,r}(\theta)=
  \Omega(\theta)^{-1}\hat{q}^{11}_{\tau,r}(\theta)=
  \frac{1}{2}\big(W^\mathrm{p}(\tau;r,\theta)+
  W^\mathrm{p}(\tau,-\theta)^\ast\big)
  \ee
  and
  \be\la{13.22}
\hat{q}^{01}_{\tau,r}(\theta)=
  -\hat{q}^{10}_{\tau,r}(\theta)=
  -\frac{i}{2}\big(W^\mathrm{p}(\tau;r,\theta)-
  W^\mathrm{p}(\tau;r,-\theta)^\ast\big)\,.
  \ee
\end{theorem}

We conclude that close to $r/\varepsilon$ in space and close to
$\tau/\varepsilon$ in time the random field $Y^j(x,t)$ is a
stationary Gaussian field. Its distribution at fixed local time
$t$ is given by $\mu^\mathrm{G}_{\tau,r}$ while in time it evolves
deterministically according $U(t)$. In this sense locally in space
and time the random field is stationary with statistics determined
through the Wigner matrix at $(r,\tau)$ and the microscopic
dynamics, compare with (\ref{13.21}), (\ref{13.22}).

\setcounter{equation}{0}
 \section{Convergence of correlation functions}

At first we introduce the matrix $q_{\tau,r}(x)$.
In Fourier space,
\beqn\la{qtaur}
\hat q_{\tau,r}(\theta)=
\sum_{\si=1}^{s}
\Pi_\si(\theta)\big({\bf M}_+^{\sigma}(\tau;r,\theta)+
{\bf M}_-^{\sigma}(\tau;r,\theta)\big)\Pi_\si(\theta),
\,\,\,\,\theta\in \T^d\setminus{\cal C}_*\,,
\eeqn
where $\Pi_\si(\theta)$ is the spectral projection
introduced in Lemma \re{lc*} $iv)$,
\beqn\la{Pi}
\ba{lll}
 {\bf M}^\sigma_{+}(\tau;r,\theta)&=&\frac{1}2\big({\bf R}^\sigma_+(\tau;r,\theta)+
C(\theta){\bf R}^\sigma_+(\tau;r,\theta)C^*(\theta)\big),\medskip\\
{\bf M}_{-}^\sigma(\tau;r,\theta)&=&\frac{1}2\big(C(\theta)
{\bf R}_{-}^\sigma(\tau;r,\theta)-{\bf R}_{-}^{\sigma}(\tau;r,\theta)C^*(\theta)\big),
\ea
\eeqn
with matrix $C(\theta)$   as in (\ref{C(theta)}) and
\beqn\la{gclimcor2}
{\bf R}^\sigma_\pm(\tau;r,\theta)&=&\frac{1}2
\big(\hat R(r+\nabla\om_\sigma(\theta)\tau,\theta)\pm\hat R(r-\nabla\om_\sigma(\theta)\tau,\theta)\big).
\eeqn

\begin{theorem} \la{the2'}
Let the conditions {\bf V1} - {\bf V2} and
{\bf E1} - {\bf E6} hold.
Then for any $r\in \R^d$, $x,y\in \Z^d$,
$\tau\neq 0$
the correlation functions of
measures $\mu^\ve_{\tau/\ve,r}$ converge to a limit,
\beqn\la{gclimcor0}
 \lim_{\ve\to 0}Q_{\ve,\tau/\ve}^{ij}([r/\ve]+x,[r/\ve]+y)
=q^{ij}_{\tau,r}(x-y).
\eeqn
\end{theorem}

We outline the strategy of the proof.
For the proof we use the cutting strategy from \ci{DKS1}
combined with some techniques from \ci{DPST},
where Theorem \ref{the2'} has proved for the case
$d=n=1$ (see \ci[Theorem 3.1]{DPST}).
Note that in \cite{DPST} it is assumed the stronger
conditions on matrix $V$ than {\bf E3}, {\bf E4}, namely, 
 $\omega(\theta)>0$, and  
the set
$$
\{\theta\in[-\pi,\pi]:\,\omega''(\theta)=\omega'''(\theta)=0\}
$$
is empty.
Under these conditions, in \ci{DPST} 
the uniform asymptotics
of the Green function is proved,
\be\la{asGrf}
\sup_{x\in\Z^d}|{\cal G}_t(x)|\le C(1+ |t|)^{-1/3}.
\ee
This bound plays an important role in the proof of \ci{DPST}.
However,
if $n > 1$, then $\omega_s$ may be non-smooth because of band crossing,
and if $d > 1$, the set where the Hessian vanishes does not consist of
isolated points.
Therefore a strong estimate as (\ref{asGrf}) is unlikely to be valid, in
general.
  To cope with such a situation, we split ${\cal G}_t(x)$
  into two summands:  ${\cal G}_t(x)={\cal G}^f_t(x)+
{\cal G}^g_t(x)$, where  ${\cal G}^f_t(x)$
has a support in the neighborhood of a a ``critical set''
${\cal C}\subset \T^d$, and ${\cal G}^g_t(x)$
vanishes in the neighborhood of ${\cal C}$.
The set ${\cal C}$ includes  all points $\theta\in \T^d$
either with a degenerate Hessian  of $\om_\sigma(\theta)$,
 or with non-smooth  $\om_\sigma(\theta)$
(see Definition \ref{calC}). We show that
 the contribution of ${\cal G}^f_t(x)$ is negligible
uniformly in $t$ (see (\ref{3.12})).
Hence, it allows us to represent correlations functions
$Q_{\ve,\tau/\ve}$ in the form:
$Q_{\ve,\tau/\ve}=Q^g_{\ve,\tau/\ve}+Q^r_{\ve,\tau/\ve}$,
such that
\beqn\nonumber
Q^g_{\ve,\tau/\ve}(x,y)=\sum\limits_{x',y'\in\Z^d}
{\cal G}^g_t(x-x')
Q_{\ve}(x',y'){\cal G}^g_t(y-y')^*.
\eeqn
For the remainder $Q^r_{\ve,\tau/\ve} =Q_{\ve,\tau/\ve}-Q^g_{\ve,\tau/\ve}$
we prove that $Q^r_{\ve,\tau/\ve}(x,y)=o(1)$
uniformly in $\tau\not=0$, $\ve>0$ and $x,y\in\Z^d$.
The last fact follows from two key observations:
i) mes${\cal C}=0$ (Lemma \ref{lc*}) and ii)
the correlation quadratic form is continuous in $\ell^2$,
see Corollary \ref{l4.1}.
Up to this point we apply the ``cutting strategy" from \ci{DKS1,DKM1}.
Finally, in Section 4.3 we prove that
$Q^g_{\ve,\tau/\ve}([r/\ve]+x,[r/\ve]+y)$
converges to a limit as $\ve\to0$, using the techniques
of \ci{DPST}.
In addition, the asymptotics of ${\cal G}^g_t(x)$,
(see Lemma \ref{l5.3}) of the form
${\cal G}^g_t(x)\sim (1+|t|)^{-d/2}$ plays the important role,
since it replaces the asymptotics (\ref{asGrf})
and also simplifies some steps of the proof of \ci{DPST}.

\subsection{Bounds for initial covariance}
\begin{definition}
By $\ell^p\equiv \ell^p(\Z^d)\otimes \R^n$, $p\ge 1$,
$n\ge 1$,
 we denote the space of sequences
$f(x)=(f_1(x),\dots,f_n(x))$ endowed with norm
$\Vert f\Vert_{p}=\Big(\sum\limits_{x\in\Z^d}|f(x)|^p\Big)^{1/p}$.
\end{definition}

\begin{lemma} \la{l4.1}
Let condition {\bf V2} hold. Then
for $i,j=0,1,$
the following bounds hold
\beqn
\sum\limits_{y\in\Z^d} |Q^{ij}_\ve(x,y)|
&\le& C<\infty\,\,\,\mbox{ for all }\,x\in\Z^d,
\nonumber\\
\sum\limits_{x\in\Z^d} |Q^{ij}_\ve(x,y)|
&\le& C<\infty\,\,\,\mbox{ for all }\,y\in\Z^d.
\nonumber
\eeqn
Here the constant $C$
does not depend on $x,y\in \Z^d$ and $\ve>0$.
\end{lemma}

\begin{cor}\la{c4.1}
Lemma \ref{l4.1} implies, by the Shur lemma,
 that for any $\Phi,\Psi\in \ell^2$
the following bound holds:
\beqn\nonumber
|\langle
Q_\ve(x,y),\Phi(x)\otimes\Psi(y)\rangle|\le
C\Vert\Phi\Vert_{2}\Vert\Psi\Vert_{2},
\eeqn
where a constant $C$ does not depend on $\ve>0$.
\end{cor}

\subsection{Stationary phase method}
Applying Fourier transform
to (\ref{CP}) we obtain
\be\la{CPF}
\dot{\hat Y}(t)=
\hat {\cal A}(\theta)\hat Y(t),\,\,\,t\in\R,
\,\,\,\,\hat Y(0)=\hat Y_0.
\ee
Here we  denote
\be\la{hA}
\hat{\cal A}(\theta)=\left(
 \begin{array}{cc}
0 & 1\\
-\hat V(\theta) & 0
\end{array}\right),\,\,\,\,\theta\in \T^d.
\ee
The solution to (\ref{CP})
admits the representation
\be\la{solGr}
Y(x,t)=\sum\limits_{y\in\Z^d}{\cal G}_t(x-y)Y_0(y),
\ee
where the Green function ${\cal G}_t(x)$
has the form
\beqn\nonumber
{\cal G}_t(x)=(2\pi)^{-d}\int\limits_{\T^d}
e^{-ix\cdot\theta}\exp\big(\hat{\cal A}(\theta)t\big)\,d\theta.
\eeqn
Note that
\beqn\la{4.5}
\hat{\cal G}_t( \theta)=
\left( \begin{array}{cc}
 \cos\Om t &~ \sin \Om t~\Om^{-1}  \\
 -\sin\Om t~\Om
&  \cos\Om t\end{array}\right),
\eeqn
where $\Om=\Om( \theta)$ is the
Hermitian matrix defined by (\ref{Omega}).
Hence,
we can rewrite ${\cal G}_t(x)$ in the form
\be\la{3.4'}
{\cal G}_t(x)=
\sum\limits_{\pm,\sigma=1}^s \int\limits_{\T^d}
e^{-ix\cdot \theta}e^{\pm i\om_\sigma(\theta)\,t}
a^\pm_\sigma(\theta)\,d\theta.
\ee
We are going to apply the stationary phase arguments
to the integral (\ref{3.4'})
 which require a smoothness
in $\theta$. Then we have to choose certain smooth
branches
of the functions $a^\pm_\sigma(\theta)$
 and $\om_\sigma(\theta)$ and cut off all singularities.
First,
we introduce the {\it critical set} as
\be\la{calC}
{\cal C} ={\cal C}_*\bigcup_{\sigma=1}^{s}{\cal C}_\sigma
 \bigcup\limits_{i=1}^{d}\bigcup\limits_{\sigma=1}^{s}\,
\Big\{\theta\in \T^d:\,
 \frac{\pa^2\omega_\sigma(\theta)}{\pa \theta_i^2}=0\Big \},
\ee
with ${\cal C}_*$ as in Lemma \ref{lc*} and
sets ${\cal C}_0$
and ${\cal C}_\sigma$ defined by (\ref{c0ck}).
Obviously
${\rm mes}\,{\cal C}=0$.
  Secondly,
fix an $\delta>0$ and choose a finite partition of unity
\beqn\nonumber
f(\theta)+ g(\theta)=1,\,\,\,\,
g(\theta)=\sum_{m=1}^M g_m(\theta),
\,\,\,\,\theta\in \T^d,
\eeqn
where $f,g_m$ are non-negative functions from
$C_0^\infty(\T^d)$, and
\be\la{fge}
\supp f\subset \{\theta\in \T^d:\,
{\rm dist}(\theta,{\cal C})<\delta\},\,\,\,
\supp g_m\subset \{\theta\in \T^d:\,
{\rm dist}(\theta,{\cal C})\ge\delta/2\}.
\ee
Then we represent ${\cal G}_t(x)$ in the form
${\cal G}_t(x)=
{\cal G}^f_t(x)+{\cal G}^g_t(x)$,
where
\beqn\la{frepecut}
{\cal G}^f_t(x)&=&(2\pi)^{-d}\int\limits_{\T^d}
e^{-ix\cdot\theta}f(\theta)\,\hat{\cal G}_t(\theta)\,d\theta,\\
\la{frepecut'}
{\cal G}^g_t(x)&=&(2\pi)^{-d}\int\limits_{\T^d}
e^{-ix\cdot\theta}g(\theta)\,\hat{\cal G}_t(\theta)\,d\theta
=\sum\limits_{\pm,\sigma=1}^s\sum\limits_{m=1}^{M} \int\limits_{\T^d}
g_m(\theta)e^{-ix\cdot\theta\pm i\om_\sigma(\theta)t}
a^\pm_\sigma(\theta)\,d\theta.
\eeqn
By Lemma \re{lc*} and the compactness arguments,
we can choose the supports of $g_m$ so small that
the eigenvalues $\om_\sigma(\theta)$
and the amplitudes $a^\pm_\sigma(\theta)$
are real-analytic functions inside
the $\supp g_m$ for every $m$. (We do not label the
functions by the index
$m$ to not overburden the notations.)
For the function ${\cal G}^f_t(x)$,
the Parseval identity, (\ref{4.5}),
and condition {\bf E6} imply
\be\la{3.12}
\Vert {\cal G}^f_t(\cdot)\Vert^2_{2}=C
\int\limits_{\T^d} |\hat{\cal G}_t(\theta)|^2|f(\theta)|^2\,d\theta
\le C\int\limits_{{\rm dist}(\theta,{\cal C})<\delta}
|\hat{\cal G}_t(\theta)|^2\,d\theta \to 0\,\,\,\mbox{as } \,\,\de\to 0,
\ee
uniformly in $t\in\R$.
For the function ${\cal G}^g_t(x)$
the following lemma holds.
\begin{lemma}\la{l5.3}
Let conditions {\bf E1} - {\bf  E4} and {\bf E6}
hold. Then
\be\la{bphi}
i)~~~~~~~~~~~~~~~~~~~~~~~~~~~~~~~~~
\sup_{x\in\Z^d}|{\cal G}^g_t(x)| \le  C~t^{-d/2}.
~~~~~~~~~~~~~~~~~~~~~~~~~~~~~~~~~~~~~~~~~
\ee
ii) For  any $p>0$ there exist
$C_p,\gamma_g>0$ such that
\be\la{conp}
|{\cal G}^g_t(x)|\le C_p(|t|+|x|+1)^{-p},\quad\quad
 |x|\ge \gamma_g t.
\ee
\end{lemma}
{\bf Proof}
Consider ${\cal G}^g_t(x)$
along each ray $x=vt$ with  arbitrary $v\in\R^d$.
By (\ref{frepecut'}), one obtains
\beqn\nonumber
{\cal G}^g_t(vt)=
\sum\limits_{m=1}^M\sum\limits_{\pm,\sigma=1}^s \int\limits_{\T^d} g_m(\theta)
e^{-i(\theta \cdot v\mp\om_\sigma(\theta))\, t}a^\pm_\sigma(\theta)\,d\theta.
\eeqn
This is a sum of oscillatory integrals with the phase
functions
$\phi^\pm_\sigma(\theta)=\theta\cdot v\mp\om_\sigma(\theta)$.
Since $\om_\sigma(\theta)$ is real-analytic,
 each function $\phi^\pm_\sigma$  has no more than a
 finite number  of  stationary points $\theta\in\supp g$,
which are solutions to the equation $v=\pm\nabla\om_\sigma(\theta)$.
The stationary points are non-degenerate for
$\theta\in\supp g_m$, by (\ref{calC}) and  (\ref{fge}), since
\beqn\nonumber
{\rm det}\Big(\frac{\pa^2 \phi^\pm_\sigma}
{\pa \theta_i\pa \theta_j}\Big)=\mp D_\sigma(\theta)\not= 0,
\,\,\,\,\,\theta\in \supp g_m.
\eeqn
Therefore, ${\cal G}^g_t(vt)={\cal O}(t^{-d/2})$
 according to the standard stationary phase method
 \ci{F, RS3}.  This implies the bounds (\ref{bphi})
in each cone $|x|\le ct$ with any finite $c$.

 Further, denote by
$\bar v_g =\max\limits_{m}\max\limits_{\sigma=1,\dots,s}
\max\limits_{\theta\in \supp g_m}|\nabla \om_\sigma(\theta)|$.
Then for $|v|>\bar v_g$
the stationary points do not exist on the $\supp g $.
Hence, the integration by parts as in \ci{RS3} yields
${\cal G}^g_t(vt)={\cal O}(t^{-p})$ for any $p>0$.
On the other hand, the integration by parts
(see (\ref{frepecut}))  implies the similar bound
${\cal G}^g_t(x)=
{\cal O}\big(\ds(t/|x|)^l\big)$ for any $l>0$.
Therefore, (\ref{conp}) follows with any $\gamma_g>\bar v_g$,
which means that the bounds (\ref{bphi}) hold everywhere.
\hfill$\bo$

\subsection{Proof of Theorem
\ref{the2'}
}
{\bf Proof}\,
{\it Step i).}
The representation (\re{solGr}) gives
\beqn\la{3.4}
Q_{\ve,t}(x,y)=
\mathbb{E}^\ve_0\big(Y(x,t)\otimes Y(y,t)\big)
=\sum\limits_{x',y'\in \Z^d}
{\cal G}_t(x\!-\!x')Q_\ve(x',y'){\cal G}_t(y\!-\!y')^*
\eeqn
for any $t\in\R^1$.
Corollary \ref{c4.1} and (\ref{3.12}) imply
\beqn\nonumber
Q_{\ve,t}(x,y)
=\sum\limits_{x',y'\in \Z^d}
{\cal G}^g_t(x\!-\!x')Q_\ve(x',y')
{\cal G}^{g}_t(y\!-\!y')^*+o(1),
\eeqn
where $o(1)\to 0$ as  $\delta\to 0$
uniformly in $t\in \R$ and $x,y\in\Z^d$.
In particular, setting $t=\tau/\ve$,
$x=[r/\ve]+l$ and $y=[r/\ve]$ we get
\beqn
Q_{\ve,\tau/\ve}([r/\ve]\!+\!l,[r/\ve])&=&
\sum\limits_{x',y'\in \Z^d}
{\cal G}^g_{\tau/\ve}([r/\ve]+l-\!x')
Q_\ve(x',y'){\cal G}^{g}_{\tau/\ve}([r/\ve]-\!y')^*+o(1)\nonumber\\
&=&\sum\limits_{x',y'\in \Z^d}
{\cal G}^g_{\tau/\ve}(l+x')
Q_\ve([r/\ve]\!-\!x',[r/\ve]\!-\!y'){\cal G}^{g}_{\tau/\ve}(y')^*
+o(1).\nonumber
\eeqn
Let $c =\gamma_g+|l|$. Then
 Lemma \ref{l5.3}, ii) and
condition {\bf V2}  imply
\beqn
Q_{\ve,\tau/\ve}([r/\ve]\!+\!l,[r/\ve])&=&
\sum\limits_{x',y'\in [-c \tau/\ve,c \tau/\ve]^d\cap\Z^d}
{\cal G}^g_{\tau/\ve}(l+\!x')
Q_\ve([r/\ve]\!-\!x',[r/\ve]\!-\!y'){\cal G}^{g}_{\tau/\ve}(y')^*\nonumber\\
&&+r_1(\ve,\tau)+o(1),\nonumber
\eeqn
where $\lim\limits_{\ve\to0}\ve^{-p}r_1(\ve,\tau)=0$
for any $p>0$
and $\tau\in\R^1$.

{\it Step ii).}
We divide the cube $[-c \tau/\ve,c \tau/\ve]^d$
onto the cubes $I_{n N_\ve}$
(see (\ref{cube})),
$$
[-c \tau/\ve,c \tau/\ve]^d
 \subset\bigcup\limits_{n\in J} I_{n N_\ve},
 $$
where $J =\{n=(n_1,\dots,n_d)\in\Z^d,\,
|n_j|\le[c \tau/(\ve N_\ve)]+1\}$.
Then
\beqn
Q_{\ve,\tau/\ve}([r/\ve]\!+\!l,[r/\ve])&=&
\sum\limits_{m,n\in J}\!\!\!
\sum\limits_{\scriptsize\ba{c}
x'\in I_{m N_\ve}\\
y'\in I_{n N_\ve}\ea}\!\!\!\!\!\!\!
{\cal G}^g_{\tau/\ve}(l+\!x')
Q_\ve([r/\ve]\!-\!x',[r/\ve]\!-\!y'){\cal G}^{g}_{\tau/\ve}(y')^*\nonumber\\
&&+r_1(\ve,\tau)+o(1)\nonumber\\
&=&\sum\limits_{m\in J}
\sum\limits_{x', y'\in I_{m N_\ve}}
{\cal G}^g_{\tau/\ve}(l+\!x')
Q_\ve([r/\ve]\!-\!x',[r/\ve]\!-\!y'){\cal G}^{g}_{\tau/\ve}(y')^*\nonumber\\
&&+r_2(\ve,\tau)+r_1(\ve,\tau)+o(1),\nonumber
\eeqn
where
\be\la{r2}
r_2(\ve,\tau) =\sum\limits_{\scriptsize\ba{c}m,n\in J,\,m\not=n\\
x'\in I_{m N_\ve},\, y'\in I_{n N_\ve}\ea}
{\cal G}^g_{\tau/\ve}(l+\!x')
Q_\ve([r/\ve]\!-\!x',[r/\ve]\!-\!y'){\cal G}^{g}_{\tau/\ve}(y')^*.
\ee
Now we prove that
\be\la{r2(ve,t)}
r_2(\ve,\tau) \to 0\quad\mbox{as }\quad \ve\to0
\ee
 for any $\tau\in\R^1$.
Indeed, we divide the sum in the RHS of (\ref{r2})
onto two sums $S_1$ and $S_2$, where
the first sum  $S_1$ is taken over
all $x'\in I_{m N_\ve}$ and $y'\in I_{n N_\ve}$
and $m,n\in J$ such that $\exists j\in\{1,\dots,d\}:|m_j-n_j|\ge2$;
the sum $S_2$ is taken over
all $x'\in I_{m N_\ve}$ and $y'\in I_{n N_\ve}$
and $m,n\in J$ such that $m\not=n$ and
$\forall j=1,\dots,d: |m_j-n_j|\le1$.
By Lemma  \ref{l5.3}, i) and condition {\bf V2},
the sum $S_1$ is estimated by
$$
 C(1+\tau/\ve)^{-d}(\tau/\ve)^d
\sum\limits_{s\in\Z^d,|s|\ge N_\ve}(1+|s|)^{-\gamma},
$$
which vanishes as $\ve \to 0$, since $N_\ve\to+\infty$
 and $\gamma >d$.
To estimate the second sum $S_2$ (the contribution
of nearest neighbors $I_{m N_\ve}$ and $I_{n N_\ve}$)
we choose a number $p>d+1$ and
divide the sum onto two sums:
$S_2=S_{21}+S_{22}$,
where the sum $S_{21}$ is taken over
all $m\in J$ and $x'\in I_{mN_\ve}$,
$n\in\{n\in J:\, n\not=m, \forall j: |m_j-n_j|\le1\}$
and $y'\in I_{nN_\ve}$ such that $|x'-y'|\ge N^{1/p}_{\ve}$
and the second sum $S_{22}$ is taken, respectively,
 over $y'$ such that $|x'-y'|\le N^{1/p}_{\ve}$.
The contribution of ``non-boundary zones''
$S_{21}$ is
$$
 C(1+\tau/\ve)^{-d}(\tau/\ve)^d
\sum\limits_{s\in\Z^d,|s|\ge N^{1/p}_\ve}(1+|s|)^{-\gamma}
$$
which vanishes as $\ve \to 0$.
The contribution of ``boundary zones'' $S_{22}$
is order of
\be\la{3.23}
 C(1+\tau/\ve)^{-d}(\tau/\ve N_{\ve})^d N_{\ve}^{1/p+d-1}
N_{\ve}^{d/p}\sim C N_{\ve}^{(d+1)/p-1}.
\ee
The number $p$ is chosen  such that $(d+1)/p-1<0$. Hence,
 (\ref{3.23})  vanishes as $\ve \to 0$
by condition {\bf V1}, ii).
The decay (\ref{r2(ve,t)}) is proved.

{\it Step iii).}
Now we can apply the condition
 {\bf V1}, i) at the points
$[r/\ve]-x', [r/\ve]-y'$ of the same cube
$ I_{[r/\ve]-m N_\ve}$ and obtain
$$
|Q_\ve([r/\ve]\!-\!x',[r/\ve]\!-\!y')-
R(\ve[r/\ve]-\ve m N_\ve, y'-x')|\le
C \min[(1+|x'-y'|)^{-\gamma},\ve N_{\ve}].
$$
Then
\beqn\la{3.24}
Q_{\ve,\tau/\ve}([r/\ve]\!+\!l,[r/\ve])&=&
\sum\limits_{m\in J}
\sum\limits_{x',y'\in I_{m N_\ve}}\!\!\!
{\cal G}^g_{\tau/\ve}(l+\!x')
R(\ve[r/\ve]-\ve m N_\ve, y'-x'){\cal G}^{g}_{\tau/\ve}(y')^*\nonumber\\
&&+r_3(\ve,\tau)+r_2(\ve,\tau)+r_1(\ve,\tau)+o(1).
\eeqn
Let us prove that
$\lim\limits_{\ve \to0}r_3(\ve,\tau)=0$
for any $\tau\in\R^1$.
Indeed, since
for fixed $x'\in I_{mN_\ve}$ the sum
$\sum\limits_{y'\in I_{m N_\ve}}
\min[(1+|x'-y'|)^{-\gamma},\ve N_{\ve}]$
is the order of $(\ve N_\ve)^{1-d/\gamma}$,
we get, by Lemma  \ref{l5.3}, i),
\beqn
|r_3(\ve,\tau)|&\le&
C \sum\limits_{m\in J}
\sum\limits_{x', y'\in I_{m N_\ve}}
\big|{\cal G}^g_{\tau/\ve}(l+\!x')\big|\,
\min[(1+|x'-y'|)^{-\gamma},\ve N_{\ve}]
\,\big|{\cal G}^{g}_{\tau/\ve}(y')^*\big|\nonumber\\
&\le& C(1+\tau/\ve)^{-d}(\tau/(\ve N_\ve))^d N_\ve^d(\ve N_\ve)^{1-d/\gamma}
\sim \ve^{(1-\beta)(1-d/\gamma)} \to0,\,\,\,\,\ve\to0,
\nonumber\eeqn
 by condition {\bf V1} ii),
since $\beta<1$ and $\gamma>d$.

{\it Step iv).}
By similar arguments, as in {\it steps i)} and {\it ii)},
the sums in the RHS of (\ref{3.24}) can be taken
over $\{y'\in\Z^d,\, m\in J,\,x'\in I_{mN_\ve}\}$.
The sum in $y'$ is a convolution which can be
expressed by the product in the Fourier transform:
 \beqn
Q_{\ve,\tau/\ve}([r/\ve]\!+\!l,[r/\ve])&=&
(2\pi)^{-2d}\sum\limits_{m\in J}
\sum\limits_{x'\in I_{m N_\ve}}
\int\limits_{\T^{2d}}e^{-i\theta\cdot l}e^{ix'\cdot(\theta'-\theta)}
\hat{\cal G}^g_{\tau/\ve}(\theta)
\hat R(\ve[r/\ve]-\ve m N_\ve, \theta')
\nonumber\\
&&\times\hat{\cal G}^{g}_{\tau/\ve}(\theta')^*\,d\theta d\theta'
+o_\tau(1)+o(1),
\nonumber
\eeqn
where $o_\tau(1) \to 0$ as $\ve\to 0$ for any $\tau\in\R^1\setminus\{0\}$.
Further, since
$I_{mN_\ve}=\{x'\in\Z^d:
(m_j-1/2)N_\ve\le x'_j<(m_j+1/2)N_\ve,\,j=1,\dots, d\}$, then
$$
\sum\limits_{x'\in I_{m N_\ve}}e^{ix'\cdot(\theta'-\theta)}=
\prod\limits_{j=1}^d\frac{F(\theta'_j-\theta_j,N_\ve,m_j)}
{e^{i(\theta'_j-\theta_j)}-1},
$$
where
$F(\theta_j,N_\ve,m_j) =
e^{i\theta_j N_\ve(m_j+1/2)}-
e^{i\theta_j N_\ve(m_j-1/2)}$.
Changing variables i) $(\theta,\theta')\to
(z,\theta')$, $z=\theta'-\theta$ and
ii) $(z,\theta')\to
(z,\theta)$, $\theta=\theta'$, one obtains
 \beqn\la{3.25}
Q_{\ve,\tau/\ve}([r/\ve]\!+\!l,[r/\ve])&=&
(2\pi)^{-2d}\sum\limits_{m\in J}
\int\limits_{[-\pi,\pi]^{2d}}e^{-i(\theta-z)\cdot l}
\prod\limits_{j=1}^d \frac{\al(z_j)F(z_j,N_\ve,m_j)}
{iz_j}\hat{\cal G}^g_{\tau/\ve}(\theta-z)\nonumber\\
&&\times
\hat R(\ve[r/\ve]-\ve m N_\ve, \theta)
\hat{\cal G}^{g}_{\tau/\ve}(\theta)^*\,d\theta dz
+o_\tau(1),
\eeqn
where
$\ds\al(z)=\frac{iz}{e^{iz}-1}$ if $z\in(-\pi,\pi)\setminus0$
and $\al(0)=1$.
Note that
$$
\hat{\cal G}^g_{t}(\theta)=g(\theta)\big(
\cos\Om(\theta)t+\sin\Om(\theta)t\, C(\theta)
\big).
$$
Hence, in the integrand in (\ref{3.25}) we have
for $t=\tau/\ve$,
\beqn\la{3.25'}
&&\hat{\cal G}^g_{t}(\theta-z)
\hat R(\ve[r/\ve]-\ve m N_\ve,\theta)\hat{\cal G}^{g}_{t}(\theta)^*\nonumber\\
&=&\sum\limits_{\sigma,\sigma'=1}^s
\Pi_\sigma(\theta-z)g(\theta-z)\big(
\cos\om_\sigma(\theta-z)t+\sin\om_\sigma(\theta-z)t\, C_\sigma(\theta-z)\big)
\nonumber\\
&&\hat R(\ve[r/\ve]-\ve m N_\ve,\theta)g(\theta)\big(
\cos\om_{\sigma'}(\theta)t+\sin\om_{\sigma'}(\theta)t\, C^*_{\sigma'}(\theta)\big)\Pi_{\sigma'}(\theta),
\eeqn
where
 $C_\sigma(\theta) =\left(\ba{cc}0&1/\om_\sigma(\theta)\\
-\om_\sigma(\theta)&0\ea\right)$.
Let us consider one of the terms in (\ref{3.25}).
The proof for the remaining terms is similar,
 \beqn\la{3.26}
I_\ve& =&(2\pi)^{-2d}\frac14
\int\limits_{[-\pi,\pi]^d}e^{-i\theta\cdot l}
e^{i \om_{\sigma'}(\theta)\tau/\ve}g(\theta)
\sum\limits_{m\in J}\Big(\int\limits_{-\pi}^{\pi}
e^{iz_d l_d} \frac{\al(z_d)F(z_d,N_\ve,m_d)}
{iz_d}\ldots\nonumber\\
&&\times\Big(\int\limits_{-\pi}^{\pi}
e^{iz_2 l_2} \frac{\al(z_2)F(z_2,N_\ve,m_2)}
{iz_2}
\Big(\int\limits_{-\pi}^\pi
e^{iz_1 l_1} \frac{\al(z_1)F(z_1,N_\ve,m_1)}
{iz_1}
e^{\pm i \om_\sigma(\theta-z)\tau/\ve}
\nonumber\\
&&\times
g(\theta-z)\Pi_\sigma(\theta-z)\hat R(\ve[r/\ve]-\ve m N_\ve,\theta)\Pi_{\sigma'}(\theta)
dz_1\Big)dz_{2}\Big)\dots dz_d\Big)d\theta.
\eeqn
Introduce
$\nu_1=\nu_1(\theta_1,\theta_2-z_2,\dots)=
\pm[\nabla_1\om_\sigma(\theta_1,\theta_2-z_2,\dots)\tau/(\ve N_\ve)]$,
$\nu_2=\nu_2(\theta_1,\theta_2,\theta_3-z_3,\dots)=
\pm[\nabla_2\om_\sigma(\theta_1,\theta_2,\theta_3-z_3,...)\tau/(\ve N_\ve)]$,
...,
$\nu_d=\nu_d(\theta)=\pm[\nabla_d\om_\sigma(\theta)\tau/(\ve N_\ve)]$.
\begin{lemma} Let condition {\bf I4} hold. Then
 \beqn\la{3.27}
I_\ve&=&(2\pi)^{-2d}\frac14
\int\limits_{[-\pi,\pi]^d}e^{-i\theta\cdot l}
e^{i\om_{\sigma'}(\theta)\tau/\ve}g(\theta)
\Big(\sum\limits_{|m_d-\nu_d|\le 2}
\int\limits_{-\pi}^\pi e^{iz_d l_d}
\frac{\al(z_d)F(z_d,N_\ve,m_d)}
{iz_d}\ldots\nonumber\\
&&\times\Big(\sum\limits_{|m_2-\nu_2|\le 2}
\int\limits_{-\pi}^\pi e^{iz_2 l_2}
\frac{\al(z_2)F(z_2,N_\ve,m_2)}
{iz_2}\Big(\sum\limits_{|m_1-\nu_1|\le 2}
\int\limits_{-\pi}^\pi e^{iz_1 l_1}
\frac{\al(z_1)F(z_1,N_\ve,m_1)}{iz_1}g(\theta\!-\!z)
\nonumber\\
&&
\times e^{\pm i \om_\sigma(\theta-z)\tau/\ve} \Pi_{\sigma}(\theta-z)\hat R(\ve[r/\ve]-\ve m N_\ve,\theta)\Pi_{\sigma'}(\theta)
dz_1\Big) dz_2\Big)\dots dz_d \Big)d\theta+o_\tau(1),
\eeqn
where $o_\tau(1)\to 0$ as $\ve\to0$
for any $\tau\in\R^1$.
\end{lemma}
{\bf Proof}. We generalize the strategy of the proof
of Proposition 3.6 from \ci{DPST},
where this assertion is proved for $d=1$.
The asymptotics (\ref{3.27}) follows from
(\ref{3.26}) if we prove that the series
over $\max_j|m_j-\nu_j|\ge 3$
vanishes as $\ve\to0$.

First, let us consider
the inner integral over $z_1$ in (\ref{3.26})
and denote it by $I_\ve(\theta,z',m)$:
\beqn\nonumber
I_\ve(\theta,z',m)=\int\limits_{-\pi}^\pi
a(\theta,z,m)\frac{e^{if_+(\theta,z,m_1)N_\ve}-
e^{if_-(\theta,z,m_1)N_\ve}}{iz_1}\,dz_1,
\eeqn
where $a(\theta,z,m)= \al(z_1)g(\theta-z)
\Pi_{\sigma}(\theta-z)\hat R(\ve[r/\ve]-\ve m N_\ve,\theta)\Pi_{\sigma'}(\theta)$,
$z'=(z_2,\dots,z_d)\in[-\pi,\pi]^{d-1}$,
$\theta\in[-\pi,\pi]^d$ and
\beqn
f_+(\theta,z,m_1)&=&z_1l_1/N_\ve
+z_1(m_1+1/2)\pm  \om_\sigma(\theta-z)\tau/(\ve N_\ve),\nonumber\\
f_-(\theta,z,m_1)&=&z_1l_1/N_\ve
+z_1(m_1-1/2)\pm  \om_\sigma(\theta-z)\tau/(\ve N_\ve).\nonumber
\eeqn
We have
$f_+(\theta,z,m_1)\Big|_{z_1=0}=f_-(\theta,z,m_1)\Big|_{z_1=0}=
\pm  \om_\sigma(\theta_1,\theta_2-z_2,\dots)\tau/(\ve N_\ve)$,
and
\beqn\nonumber
\nabla_1f_\pm(\theta,z,m_1)\big|_{z_1=0}&=& l_1/N_\ve
+m_1\pm 1/2-\nu_1.
\eeqn
Hence, $\nabla_1f_\pm(\theta,z,m_1)\Big|_{z_1=0}\not=0$
for $|m_1-\nu_1|\ge 3$.
Indeed, we can admit that $|l_1/N_\ve|\le1$
since $N_\ve\to\infty$
as $\ve\to0$, and the number $l_1\in\Z$ is fixed.
Further, we apply to $I_\ve(\theta,z',m)$
the limit of Lemma 3.7 from \ci{DPST},
\beqn\la{3.28}
&&\lim_{\ve\to0}
\Big[I_\ve(\theta,z',m)-
\pi e^{\pm i\om_\sigma(\theta_1,\theta_2-z_2,...)\tau/\ve}
a(\theta,(0,z'),m)
\nonumber\\
&&\times\Big({\rm sgn}\nabla_1f_+
(\theta,z,m_1)|_{z_1=0}-
{\rm sgn}\nabla_1f_-(\theta,z,m_1)|_{z_1=0}\Big)
\Big]=0.
\eeqn
Moreover, we obtain that
$I_\ve(\theta,z',m)\to0$ as $\ve\to0$,
uniformly in $\theta\in[-\pi,\pi]^d$ and
$z'\in[-\pi,\pi]^{d-1}$,
since
${\rm sgn}\nabla_1f_+(\theta,z,m_1)\Big|_{z_1=0}=
{\rm sgn}\nabla_1f_-(\theta,z,m_1)\Big|_{z_1=0}$.
We proceed by induction for
each inner integral over $z_2,\dots,z_d$
and obtain that the integrals with
$\max_j|m_j-\nu_j|\ge3$  vanish as $\ve\to0$.
Further, we have to prove that  the series
over $\max_j|m_j-\nu_j|\ge3$ also vanish.
This follows from two facts:
i) the function  $a(\theta,z,m)$ has a structure of
$f(\theta,z)\hat R(\ve[r/\ve]-\ve m N_\ve,\theta)$
with a smooth function $f$,
and ii) $\hat R(r,\theta)$ satisfies condition {\bf I4}.
More exactly it is proved for the case $d=1$
in \ci{DPST}. The proof
 admits generalization to the case $d>1$,
here we omit the detailed computations.

{\it Step v)} The next step is to prove that
 \beqn\la{3.29}
I_\ve&=&\frac{(2\pi)^{-2d}}4
\int\limits_{[-\pi,\pi]^{d}}
e^{-i\theta\cdot l+i \om_l(\theta)\tau/\ve}
g(\theta)\Pi_{\sigma}(\theta)
\hat R(r\mp\nabla\om_{\sigma'}(\theta)\tau,\theta)
\Pi_{\sigma'}(\theta)\Big(\int\limits_{-\pi}^\pi
e^{iz_d l_d+i\nu_d N_\ve z_d}
\al(z_d)\nonumber\\
&&\frac{e^{i5/2 N_\ve z_d}-
e^{-i5/2 N_\ve z_d}}{iz_d}\ldots
\Big(\int\limits_{-\pi}^\pi
e^{iz_2 l_2+i\nu_2 N_\ve z_2}
\al(z_2)\frac{e^{i5/2 N_\ve z_2}-
e^{-i5/2 N_\ve z_2}}{iz_2}\nonumber\\
&&\Big(\int\limits_{-\pi}^\pi
e^{iz_1 l_1+i\nu_1 N_\ve z_1\pm i \om_\sigma(\theta-z)
\tau/\ve}
\al(z_1)\frac{e^{i5/2 N_\ve z_1}-
e^{-i5/2 N_\ve z_1}}{iz_1} g(\theta-z)
\,dz_1\Big) dz_2\Big)\dots dz_d\Big)d\theta\nonumber\\
&&+o_\tau(1),
\eeqn
where $o_\tau(1)\to 0$ as $\ve\to0$
for any $\tau\in\R^1$.
It follows from (\ref{3.27}) and
the formula
$\sum_{|m_j-\nu_j|\le2}F(z_j,N_\ve,m_j)=
e^{i\nu_j N_\ve z_j}(e^{i5/2 N_\ve z_j}-e^{-i5/2 N_\ve z_j})$.
 Formula (\ref{3.29}) is proved
in Lemma 3.8 from \ci{DPST} for the case $d=1$.
The proof is based on the condition {\bf I4}
for function $\hat R$ and admits extension to the case $d>1$.

Further, we apply (\ref{3.28})
to the inner integrals from
the RHS of  (\ref{3.29})  and obtain,
 for the inner integral over $z_1$ (denote it by $I_\ve
(\theta,z')$, where $z'=(z_2,\dots,z_d)$),
\beqn
\lim_{\ve\to0}
\left[I_\ve(\theta,z')-
2\pi e^{\pm i\om_\sigma(\theta_1,\theta_2-z_2,...)\tau/\ve}
g(\theta_1,\theta_2-z_2,\dots,\theta_d-z_d)\right]=0,
\eeqn
since in this case
$I_\ve(\theta,z')=\int_{\T^1}\al(z_1)g(\theta-z)\big(
\exp{(if_+(\theta,z)N_\ve)}
-\exp{(if_-(\theta,z)N_\ve)}\big)/(iz_1)\,dz_1$
with
\beqn
f_+(\theta,z)&=&z_1l_1/N_\ve
+z_15/2+\nu_1 z_1\pm  \om_\sigma(\theta-z)\tau/(\ve N_\ve),\nonumber\\
f_-(\theta,z)&=&z_1l_1/N_\ve
-z_1 5/2+\nu_1 z_1\pm  \om_\sigma(\theta-z)\tau/(\ve N_\ve),\nonumber
\eeqn
and
${\rm sgn}\nabla_1f_\pm(\theta,z)\Big|_{z_1=0}=
{\rm sgn}(l_1/N_\ve\pm5/2)=\pm1$
for fixed $l_1\in\Z$ and small enough $\ve>0$.
Finally,  we obtain
 \beqn\la{3.30}
I_\ve=\frac{(2\pi)^{-d}}4
\int\limits_{\T^{d}}e^{-i\theta\cdot l}
e^{i \big(\om_{\sigma'}(\theta)\pm \om_\sigma(\theta)\big)\tau/\ve}
g(\theta)\Pi_{\sigma}(\theta)
\hat R(r\mp\nabla\om_\sigma(\theta)\tau,\theta)\Pi_{\sigma'}(\theta)\,d\theta
+o_\tau(1),
\eeqn
where $o_\tau(1)\to 0$ as $\ve\to0$
for any $\tau\not=0$.

{\it Step vi)}
Note that the identities
$\om_\sigma(\theta)\pm\om_{\sigma'}(\theta)\equiv{\rm const}_\pm$  in the exponent
(see (\ref{3.30}))
with the ${\rm const}_\pm\ne 0$ are
impossible by the condition {\bf E5}.
Furthermore,
the oscillatory integrals with $\om_\sigma(\theta)\pm\om_{\sigma'}(\theta)\not\equiv{\rm const}_\pm$
vanish as $\ve\to0$ by the condition {\bf I1}
and the Lebesgue-Riemann theorem.
Hence,
only the integrals with
$\om_\sigma(\theta)-\om_{\sigma'}(\theta)\equiv 0$
contribute to the integral  (\ref{3.30})
since  $\om_\sigma(\theta)+\om_{\sigma'}(\theta)\equiv 0$ would imply
$\om_\sigma(\theta)\equiv\om_{\sigma'}(\theta)\equiv 0$ which
is impossible by  {\bf E4}.
We return to formula (\ref{3.25}) and applying (\ref{3.25'})
one obtains formulas (\ref{qtaur}).
\hfill$\bo$

\setcounter{equation}{0}
\section{Proof of Theorems
\ref{the2}
and
\ref{the3}
}
\subsection{Convergence of Wigner matrices}
{\bf Proof of Theorem \ref{the2}.}
Theorem \ref{the2'} implies that
 for any $r\in\R^d$,
$\tau\not=0$ and $y\in (2\Z)^d$ the following convergence holds,
\beqn\la{5.1}
\lim_{\ve\to0}
\mathbb{E}^\varepsilon_{\tau/\varepsilon}
\big(a([r/\ve]+y/2)^\ast \otimes a([r/\ve]-y/2)\big)
={\cal W}^\mathrm{p}(\tau;r,y),
\eeqn
where in the Fourier space one has
\beqn\la{5.2}
\hat {\cal W}^\mathrm{p}(\tau;r,\theta)&=&
 \frac{1}{2}\Big(\Omega^{1/2}\hat q^{00}_{\tau,r}(\theta)\Omega^{1/2}+
 \Omega^{-1/2}\hat q^{11}_{\tau,r}(\theta)\Omega^{-1/2}
\nonumber\\&&
+i\Omega^{1/2}\hat q^{01}_{\tau,r}(\theta)\Omega^{-1/2}-i
 \Omega^{-1/2}\hat q^{10}_{\tau,r}(\theta)\Omega^{1/2}\Big)
\nonumber\\
&=&W^\mathrm{p}(\tau;r,\theta),
\eeqn
by formulas (\ref{3.12'}), (\ref{3.16a}) and
(\ref{qtaur})--(\ref{gclimcor2}).
Then convergence (\ref{3.17}) follows from (\ref{5.1}),
(\ref{5.2}) and Lemma \ref{lcom}.

\begin{lemma}\la{lcom}
Let conditions {\bf V2} and {\bf E1} - {\bf E3}, {\bf E6} hold and
$\al<-d/2$. Then
\beqn\la{sup}
\sup\limits_{\ve,t\in\R} \sup\limits_{x,y\in \Z^d}
\Vert Q_{\ve,t}(x,y)\Vert\le C<\infty.
\eeqn
\end{lemma}
{\bf Proof}
Applying (\re{3.4}) one has
\beqn\nonumber
Q^{ij}_{\ve,t}(x,y)=
\mathbb{E}^\ve_0\big(Y^i(x,t)\otimes Y^j(y,t)\big)
=\langle Q_\ve(x',y'), \Phi^i_{x}(x',t)\otimes
\Phi^j_{y}(y',t)\rangle,
\eeqn
where
$$
\Phi^i_{x}(x',t) =\big(
{\cal G}^{i0}_t(x-x'),{\cal G}^{i1}_t(x-x')\big),\,\,
\,\,\,x'\in\Z^d,\,\,\,\,\,\,i=0,1.
$$
Then the Parseval identity, (\ref{4.5}) and condition {\bf E6}
 imply
$$
\Vert\Phi^i_{x}(\cdot,t)\Vert^2_{2}= (2\pi)^{-d}
\int\limits_{\T^d}
|\hat\Phi^i_{x}(\theta,t)|^2\,d\theta
=(2\pi)^{-d}\int\limits_{\T^d}
\big(
|\hat{\cal G}^{i0}_t(\theta)|^2
+|\hat{\cal G}^{i1}_t(\theta)|^2\big)
\,d\theta\le C_0<\infty.
$$
Then Corollary \ref{c4.1} gives
\beqn\nonumber
|Q^{ij}_{\ve,t}(x,y)|=
|\langle Q_\ve(x',y'), \Phi^i_{x}(x',t)\otimes
\Phi^j_{y}(y',t)\rangle|
\le C\Vert\Phi^i_{x}(\cdot,t)\Vert_{2}\,
\Vert\Phi^j_{y}(\cdot,t)\Vert_{2}
\le C_1<\infty,
\eeqn
where the constant $ C_1$  does  not depend on
$x,y\in\Z^d$, $t\in\R$ and $\ve>0$.
\hfill$\bo$

\subsection{Weak convergence of measures
$\mu^\ve_{\tau/\ve,r}$ as $\ve\to 0$}

Theorem  \ref{the3} follows
 from Propositions  \re{l2.1} and \re{l2.2}.
Proposition \ref{l2.1} ensures the existence of the limit measures
of the family $\{\mu^\ve_{\tau/\ve,r},\,\ve>0\}$,
while Proposition \ref{l2.2} provides the
 uniqueness.
\begin{pro}\la{l2.1}
Let conditions {\bf V2}  and {\bf E1} - {\bf E3}, {\bf E6} hold.
Then for any
$r\in \R^d$, $\tau\not=0$, the family of
measures $\{\mu^\ve_{\tau/\ve,r},\,\ve>0\}$
 is weakly compact in   ${\cal H}_\al$ with any
 $\al<-d/2$, and the following bound holds,
\be\la{p3.1}
\sup\limits_{\ve\ge 0}
 \int \Vert Y_0\Vert^2_\al\,\mu^\ve_{\tau/\ve,r}(dY_0) <\infty.
\ee
 \end{pro}
{\bf Proof}\,
 Definition (\re{d1.1})  implies
\beqn\nonumber
&&\int\Vert Y_0\Vert^2_{\al}\mu^{\ve}_{\tau/\ve,r}(dY_0)=
\mathbb{E}^\ve_0
\big(\Vert T_{-[r/\ve]} U(\tau/\ve)Y_0\Vert^2_\al\big)
\nonumber\\
&&=
\sum\limits_{x\in \Z^d}
(1+|x|^2)^\al
\Big({\rm tr}\,Q_{\ve,\tau/\ve}^{00}([r/\ve]+x,[r/\ve]+x)+
{\rm tr}\,Q_{\ve,\tau/\ve}^{11}([r/\ve]+x,[r/\ve]+x)\Big).\nonumber
\eeqn
Since  $\al<-d/2$, (\ref{p3.1}) follows from
the bound (\ref{sup}).
Now the compactness of the
measures  family  $\{\mu_t,\,t\in \R\}$ follows
from the bound (\re{p3.1})
by the Prokhorov Theorem  \cite[Lemma II.3.1]{VF}
using the method  of  \ci[Theorem XII.5.2]{VF},
since the embedding ${\cal H}_\al \subset{\cal H}_\beta$
is compact if $\al>\beta$.
\hfill$\bo$\\

Denote by ${\cal Q}_{\tau,r}$ the
quadratic form with the matrix  kernel
$(q^{ij}_{\tau,r}(x-y))_{i,j=0,1}$,
\be\la{qpp'}
{\cal Q}_{\tau,r}(\Psi,\Psi)=\sum\limits_{i,j=0,1}~
\sum\limits_{x,y\in\Z^d}
\big( q^{ij}_{\tau,r}(x-y),\Psi^i(x)\otimes\Psi^j(y)\big),
\,\,\,\,\Psi\in{\cal D}.
\ee
 \begin{pro}\la{l2.2}
Let conditions {\bf V1} - {\bf V4}  and {\bf E1} - {\bf E6} hold.
Then for any $r\in \R^d$, $\tau\not=0$
and $\Psi\in {\cal D}$,
 \be\la{2.6}
 \lim_{\varepsilon\to 0}
\int \exp(i\langle Y,\Psi\rangle)
\mu^\ve_{\tau/\ve,r}(dY)
 = \exp\big\{-\fr{1}{2}{\cal Q}_{\tau,r} (\Psi,\Psi)\big\}\,.
 \ee
 \end{pro}

Proposition \ref{l2.2} is proved in Sections 6 - 9.

\setcounter{equation}{0}
 \section{Convergence of characteristic functionals}
To prove Theorem \ref{the3},
it remains to check Proposition \ref{l2.2}.
Let us rewrite  (\re{2.6}) as
\be\la{*}
\hat \mu^\ve_{\tau/\ve,r}(\Psi)= \mathbb{E}^\ve_0
\big(\exp \{i\langle T_{-[r/\ve]}U(\tau/\ve)Y_0,\Psi\rangle\}\big)
\to \hat \mu^G_{\tau,r}(\Psi),\,\,\,\,
\ve\to0.
\ee
We will prove it in Sections 8, 9.
In   this section   we evaluate
$\langle T_{-[r/\ve]}U(t)Y_0,\Psi\rangle$,
$t\in\R$,
by using the following duality arguments.

\subsection{Duality arguments}
Remember that $Y_0\in {\cal H}_\al$
with $\al<-d/2$.
For  $t\in\R$ introduce a `formal adjoint' operator
$U'(t)$ from  space ${\cal D}$ to ${\cal H}_{-\al}$:
\be\la{def}
\langle Y,U'(t)\Psi\rangle =
\langle U(t)Y,\Psi\rangle ,\,\,\,
\Psi\in {\cal D},
\,\,\, Y\in {\cal H}_\al.
\ee
Let us denote by
$\Phi_r(\cdot,t)=U'(t)T_{[r/\ve]}\Psi$.
Then  using (\ref{def}) we obtain
\be\la{defY}
\langle T_{-[r/\ve]}U(t)Y_0,\Psi\rangle
=\langle Y_0,\Phi_r(\cdot,t)\rangle,
\,\,\,\,t\in\R,\,\,\,\,\ve>0,\,\,\,\,\,r\in\R^d.
\ee
The adjoint group $U'(t)$ admits
the following convenient description.
Lemma \re{ldu} below displays that
the action of group $U'(t)$ coincides with the action
of  $U(t)$, up to the order of the components.
\begin{lemma}\la{ldu}
For $\Psi=(\Psi^0,\Psi^1)\in {\cal D}$ we have
\be\la{UP}
\Phi(\cdot,t) =U'(t)\Psi= (\dot\psi(\cdot,t),\psi(\cdot,t)),
\ee
where
 $\psi(x,t)$ is the solution of Eqn (\ref{1.1'})
 with the initial  data
$(u_0,v_0)=(\Psi^1,\Psi^0)$.
\end{lemma}

The lemma allows us to construct
the oscillatory integral representation for
$\Phi_r(x,t)$.
Namely,  (\ref{UP})   implies that
in Fourier representation for
$\Phi_r(\cdot,t)=U'(t)T_{[r/\ve]}\Psi$  we have
$$
\dot{\hat \Phi}_r(\theta,t)
=\hat{\cal A}^*(\theta)\hat \Phi_r(\theta,t),\quad
\hat \Phi(t,t)=\hat{\cal G}^*_t(\theta)e^{i[r/\ve]\cdot\theta}
\hat\Psi(\theta),
$$
where
\beqn\nonumber
 \hat{\cal A}^*(\theta)=
\left( \begin{array}{cc}
0&-\hat V(\theta)\\
1&0
\end{array}   \right),\quad
\hat{\cal G}^*_t(\theta)=
\left(    \begin{array}{cc} {\rm cos}~
\Omega(\theta) t & -\Omega(\theta)~{\rm sin}~\Omega(\theta) t \\
\Om^{-1}(\theta)\sin \Omega(\theta) t
 & {\rm cos}~\Omega(\theta)t\end{array} \right).
\eeqn
 Therefore
\be\la{frep'}
\Phi_r(x,t)=(2\pi)^{-d}
\int\limits_{\T^d} e^{-i\theta\cdot x}
\hat{\cal G}^*_t(\theta)e^{i[r/\ve]\cdot\theta} \hat\Psi(\theta)\,d\theta,
\,\,\,x\in\Z^d.
\ee
\begin{definition}\la{dC}
${\cal D}^0 =\{\Psi\in{\cal D}:
\hat\Psi(\theta)=0\quad \mbox{\rm in a neighborhood of}
\quad{\cal C}\}$.
\end{definition}

From (\ref{frep'}) we obtain
\begin{lemma}\la{l5.3'}
For any fixed $\Psi \in {\cal D}^0$
the following bounds hold:

i) $|\Phi_r(x,t)| \le  C~t^{-d/2}$, $x\in\Z^d$.

ii) For  any $p>0$ there exist
$C_p,\gamma_g >0$ such that
$|\Phi_r(x,t)|\le C_p(|t|+|x|+1)^{-p}$,
$|x|\ge \gamma_g t$.
\end{lemma}
This lemma follows from  Lemma \ref{l5.3}
and  the definition of ${\cal D}^0$.

\subsection{Equicontinuity  of characteristic functionals}
Let us show that we can restrict ourselves to
$\Psi\in {\cal D}^0$.

\begin{lemma}
The convergence (\ref{*}) it suffices to prove for $\Psi\in{\cal D}^0$ only.
\end{lemma}
{\bf Proof.} {\it Step i).}
For simplicity, let us put $t =\tau/\ve$.
Denote by
$$
{\cal Q}_{\ve,t,r}(\Psi,\Psi) =
\ds\int|\langle Y_0,\Psi\rangle|^2\,d\mu^\ve_{t,r}(dY_0).
$$
We prove at first that
\be\la{cub}
\sup\limits_{\ve>0,t\in\R,r\in\R^d} |{\cal Q}_{\ve,t,r}(\Psi,\Psi)|
\le C\Vert\Psi\Vert^2_{2},\quad \Psi\in{\cal D}.
\ee
Indeed, by (\ref{defY}) we have
$$
{\cal Q}_{\ve,t,r}(\Psi,\Psi)=
\mathbb{E}^\ve_0\big(|\langle T_{-[r/\ve]}U(t)Y_0,\Psi\rangle|^2\big)
=\langle Q_{\ve}(x,y), \Phi_r(x,t)\otimes\Phi_r(y,t) \rangle.
$$
So, by Corollary \ref{c4.1} we obtain
$$
\sup\limits_{\ve>0,t\in\R,r\in\R^d}
|{\cal Q}_{\ve,t,r}(\Psi,\Psi)|\le
C\sup\limits_{t\in\R,r\in\R^d}
\Vert\Phi_r(\cdot,t)\Vert^2_{2}.
$$
Finally, by the Parseval identity and condition {\bf E6}, we get
$$
\Vert\Phi_r(\cdot,t)\Vert^2_{2}=
(2\pi)^{-d}\int\limits_{\T^d}
\Vert {\cal G}_{t}^*(\theta)\Vert|\hat\Psi(\theta)|^2\,d\theta
\le C\Vert\Psi\Vert^2_{2}.
$$
The bound (\ref{cub}) is proved.

{\it Step ii).}
By the Cauchy-Schwarz inequality,
$$
\ba{rcl}
|\hat\mu^\ve_{t,r}(\Psi_1)-\hat\mu^\ve_{t,r}(\Psi_2)|
&=& \Big|\ds\int \Big( e^{i\langle Y,\Psi_1 \rangle}-
e^{i\langle Y,\Psi_2 \rangle}\Big)\mu^\ve_{t,r}(dY)\Big|
\le \ds\int \Big|e^{i\langle Y,\Psi_1-\Psi_2 \rangle}-1\Big|
\mu^\ve_{t,r}(dY)\\~\\
&\le&
\ds\int |\langle Y,\Psi_1-\Psi_2 \rangle|\mu^\ve_{t,r}(dY)

\le \Big(\ds\int |\langle Y,\Psi_1-\Psi_2 \rangle|^2
\mu^\ve_{t,r}(dY)\Big)^{1/2}\\
~\\&=&
\big({\cal Q}_{\ve,t,r}(\Psi_1-\Psi_2, \Psi_1-\Psi_2)\big)^{1/2}
\le C\Vert\Psi_1-\Psi_2 \Vert_{2},
\ea
$$
where a constant $C$ does not depend on $\ve>0$,
$t\in\R^1$ and $r\in\R^d$.
Hence, the characteristic functionals
$\hat\mu^\ve_{t/\ve,r}(\Psi)$, $t\in\R$, $\ve>0$, $r\in\R^d$,
 are equicontinuous in the space ${\cal D}$
endowed with the norm $\ell^2$.
In the turn, the set ${\cal D}^0$ is dense in this space.
\hfill$\bo$

\setcounter{equation}{0}
\section{Bernstein's `rooms-corridors' partition}

Let us introduce a `room-corridor'  partition of the
ball $\{x\in\Z^d:~|x|\le\gamma_g  t\}$
with $\gamma_g$ from Lemma \ref{l5.3'} ii).
For $t>0$  we choose below
$\De_t,\rho_t\in \N$ (we will specify the
asymptotical relations between $t$, $\De_t$ and  $\rho_t$).
Let us set $h_t=\De_t+\rho_t$ and
\be\la{rom}
a^j=jh,\,\,\,b^j=a^j+\De_t,\,\,\,
j\in\Z,\,\,\,\,\,\,n_t=[\gamma_g t/h_t].
\ee
We call the slabs $R_t^j=\{x\in\Z^d,\, |x|\le n_t h_t:~a^j\le x_d< b^j\}$
the `rooms',
$C_t^j=\{x\in\Z^d,\,|x|\le
n_t h_t:~b^j\le x_d<  a^{j+1}\}$  the `corridors'
and $L_t=\{x\in\Z^d,\,|x|> n_t h_t\}$ the 'tails'.
Here  $x=(x_1,\dots,x_d)$,
$\De_t$ is the width of a room, and
$\rho_t$ is that of a corridor.
Let us denote  by
 $\chi_t^j$ the indicator of the room $R_t^j$,
 $\xi_t^j$ that of the corridor $C_t^j$, and
$\eta_t$ that of the tail $L_t$.
Then
\beqn\nonumber
{\sum}_j
[\chi_t^j(x)+\xi_t^j(x)]+ \eta_t(x)=1,\,\,\,x\in\Z^d,
\eeqn
where the sum ${\sum}_j$ stands for
$\sum\limits_{j=-n_t}^{n_t-1}$.
Hence we get the following  Bernstein's type representation:
\be\la{res}
\langle Y_0,\Phi_r(\cdot,t)\rangle = {\sum}_j
[\langle Y_0,\chi_t^j\Phi_r(\cdot,t)\rangle +
\langle Y_0,\xi_t^j\Phi_r(\cdot,t)\rangle ]+
\langle Y_0,\eta_t\Phi_r(\cdot,t)\rangle ).
\ee
Let us introduce the
random variables
 $ r_{t}^j$, $ c_{t}^j$, $l_{t}$ by
\be\la{100}
r_{t}^j= \langle Y_0,\chi_t^j\Phi_r(\cdot,t)\rangle,~~
c_{t}^j= \langle Y_0,\xi_t^j\Phi_r(\cdot,t)\rangle,
\,\,\,l_{t}= \langle Y_0,\eta_t\Phi_r(\cdot,t)\rangle.
\ee
Then  (\ref{res}) becomes
\be\la{razli}
\langle Y_0,\Phi_r(\cdot,t)\rangle =
{\sum}_j
(r_{t}^j+c_{t}^j)+l_{t}.
\ee
\begin{lemma}  \la{l5.1}
    Let  conditions {\bf V1} - {\bf V2} hold and $\Psi\in{\cal  D}^0$.
The following bounds hold for $t>1$:
\beqn
\mathbb{E}^\ve_0\big(|r^j_{t}|^2\big)&\le&  
C(\Psi)~\De_t/ t,\,\,\,\forall j,\la{106}\\
\mathbb{E}^\ve_0\big(|c^j_{t}|^2\big)&\le& 
C(\Psi)~\rho_t/ t,\,\,\,\forall j,\la{106''}\\
\mathbb{E}^\ve_0\big(|l_{t}|^2\big)&\le& 
C_p(\Psi)~(1+t)^{-p},\,\,\,\,\forall p>0.\la{106'''}
\eeqn
\end{lemma}
{\bf Proof}
The bound (\ref{106'''}) follows from (\ref{conp}).
We discuss  (\ref{106}), and (\ref{106''})
can be done in a similar way.
Let us express $\mathbb{E}^\ve_0\big(|r_t^j|^2\big)$ in the correlation matrices.
Definition (\ref{100})  implies by the Fubini Theorem
that
 \be\la{100rq}
\mathbb{E}^\ve_0\big(|r_{t}^j|^2\big)= \langle Q_\ve(x,y), \chi_t^j(x)
\Phi_r(x,t)\otimes\chi_t^j(y)\Phi_r(y,t) \rangle.
\ee
 According to (\ref{conp}) and (\ref{bphi}), Eqn (\ref{100rq})
 implies that
\beqn\la{er}
\mathbb{E}^\ve_0\big(|r_{t}^j|^2\big)&\le&
Ct^{-d}\sum\limits_{x,y}
\chi_t^j(x)\Vert  Q_\ve(x,y)\Vert \nonumber\\
&=&Ct^{-d}\sum\limits_{x}
\chi_t^j(x) \sum\limits_{z}
 \Vert Q_\ve(x,y)\Vert\le C \De_t/t,
\eeqn
where $\Vert Q_\ve(x,y)\Vert $ stands for the norm of a matrix
$\left(Q_\ve^{ij}(x,y)\right)$.
Therefore,  (\ref{er}) follows from Lemma \ref{l4.1}.
\hfill$\bo$

\setcounter{equation}{0}
\section{ Ibragimov-Linnik Central Limit Theorem}
In this section
we prove the convergence  (\ref{*}).
As was said, we use a version of the Central Limit Theorem
developed by Ibragimov and Linnik.
If  ${\cal Q}_{\tau,r}(\Psi,\Psi)=0$,
 the convergence (\ref{*}) is obvious.
Indeed, then,
\beqn\nonumber
&&\left| \mathbb{E}^\ve_0\big(
\exp\{i \langle Y_0,\Phi_r(\cdot,\tau/\ve)\rangle\}\big) -
 \hat \mu^G_{\tau,r}(\Psi)\right|
= \mathbb{E}^\ve_0\big(
|\exp\{i\langle Y_0,\Phi_r(\cdot,\tau/\ve)\rangle\}-1|\big)
\nonumber\\
&\le& \mathbb{E}^\ve_0\big(|\langle Y_0,\Phi_r(\cdot,\tau/\ve)\rangle|\big)
\le
\left(\mathbb{E}^\ve_0\big(|\langle Y_0,\Phi_r(\cdot,\tau/\ve)\rangle|^2
\big)\right)^{1/2}
\nonumber\\
&=&\big(\langle Q_\ve(x,y),\Phi_r(x,\tau/\ve)\otimes
\Phi_r(y,\tau/\ve)\rangle\big)^{1/2}
=
\big({\cal Q}_{\ve,\tau/\ve,r}(\Psi,\Psi)\big)^{1/2},
\eeqn
where ${\cal Q}_{\ve,\tau/\ve,r}(\Psi,\Psi)\to
{\cal Q}_{\tau,r}(\Psi,\Psi)=0$, $\ve\to0$.
Therefore,  (\ref{*}) follows from Theorem \ref{the2}.
Thus, we may assume that for a given $\Psi\in{\cal D}^0$,
\be\la{5.*}
{\cal Q}_{\tau,r}(\Psi,\Psi)\not=0.
\ee
Let us choose  $0<\de<1$ and
\be\la{rN}
\rho_t\sim t^{1-\delta},
~~~\De_t\sim\fr t{\log t},~~~~\,\,\,t\to\infty.
\ee
\begin{lemma}\la{r}
The following limit holds,
\be\la{7.15'}
n_t\Big[\Bigl(\frac{\rho_t}{t}\Bigr)^{1/2}+
(1+\rho_t)^{-\kappa}\Big]+
n_t^2\frac{\rho_t}{t}\to 0 ,\quad t\to\infty,
\ee
where a constant $\kappa>0$.
\end{lemma}
Indeed,  (\ref{rN}) implies that
$h_t=\rho_t+\De_t\sim \ds\frac{t}{\log t}$,
 $t\to\infty$.
Therefore, $n_t\sim\ds\frac{t}{h_t}\sim\log t$.
Then  (\ref{7.15'}) follows by  (\ref{rN}).
\hfill$\bo$\\

For simplicity, we put $t =\tau/\ve$.
By the triangle inequality,
\beqn
\Big|\mathbb{E}^\ve_0\big(\exp\{i \langle Y_0,\Phi_r(\cdot,t)\rangle\}
\big)
-\hat \mu^G_{\tau,r}(\Psi)\Big|
&\le &
\Big|\mathbb{E}^\ve_0\big(\exp\{i \langle Y_0,\Phi_r(\cdot,t)\rangle\}\big)
-
\mathbb{E}^\ve_0\big(\exp\{i{\sum}_j r_{t}^j\}\big)\Big|
\nonumber\\
&&\hspace{-12pt}+\Big|\exp\big\{-\frac{1}{2}{\sum}_j 
\mathbb{E}^\ve_0\big(|r_{t}^j|^2\big)\big\}
 \!-\!
\exp\big\{-\frac{1}{2} {\cal Q}_{\tau,r}(\Psi, \Psi)\big\}\Big|
 \nonumber\\
&&\hspace{-12pt}+ \Big|\mathbb{E}^\ve_0 \big(\exp\{i{\sum}_j r_{t}^j\}\big) \!-\!
\exp\big\{-\frac{1}{2}{\sum}_j 
\mathbb{E}^\ve_0\big(|r_{t}^j|^2\big)\big\}\Big|\nonumber\\
&=& I_1+I_2+I_3. \la{4.99}
\eeqn
We are going to   show  that all summands
$I_1$, $I_2$, $I_3$  tend to zero
 as  $t\to\infty$.\\
{\it Step (i)} Eqn (\ref{razli}) implies
\beqn\la{101}
I_1&=&\Big|\mathbb{E}^\ve_0\big(\exp\{i{\sum}_j r^j_{t} \}
\big(\exp\{i{\sum}_j c^j_{t}+il_{t}\}-1\big)\big)\Big|
\nonumber\\
&\le&
 {\sum}_j \mathbb{E}^\ve_0\big(|c^j_t|\big)+
\mathbb{E}^\ve_0\big(|l_{t}|\big)
\le{\sum}_j\Big(\mathbb{E}^\ve_0\big(|c^j_t|^2\big)\Big)^{1/2}+
\Big(\mathbb{E}^\ve_0\big(|l_{t}|^2\big)\Big)^{1/2}.
\eeqn
From (\ref{101}), (\ref{106''}), (\ref{106'''})
  and (\ref{7.15'}) we obtain that
\beqn\nonumber
I_1\le C n_t(\rho_t/t)^{1/2}+C_p t^{-p} \to 0,~~t\to \infty.
\eeqn
{\it Step (ii)} By the triangle inequality,
\beqn
I_2&\le& \frac{1}{2}
\Big|{\sum}_j \mathbb{E}^\ve_0(|r_t^j|^2)- 
{\cal Q}_{\tau,r}(\Psi,\Psi)\Big|
\le \frac{1}{2}\,
\Big|{\cal Q}_{\ve,t,r}(\Psi, \Psi)-
{\cal Q}_{\tau,r}(\Psi,\Psi)\Big|
\nonumber\\
&&+ \frac{1}{2}\, \Big|\mathbb{E}^\ve_0\big(\big({\sum}_j r_t^j\big)^2\big)
-{\sum}_j \mathbb{E}^\ve_0\big(|r_t^j|^2\big)\Big| +
 \frac{1}{2}\, \Big|\mathbb{E}^\ve_0\big(\big({\sum}_j r_t^j\big)^2\big)
-{\cal Q}_{\ve,t,r}(\Psi, \Psi)\Big|\nonumber\\
&=& I_{21} +I_{22}+I_{23}\la{104},
\eeqn
where ${\cal Q}_{\ve,t,r}$ is the quadratic form with
the  matrix kernel $Q_{\ve,t,r}^{ij}(x,y)$.
Theorem \ref{the2} implies that  $I_{21}\big|_{t=\tau/\ve}\to 0$
as $\ve\to0$.
As for  $I_{22}$,  we first obtain that
\be\la{i22}
I_{22}\le \sum\limits_{j<l} 
\left|\mathbb{E}^\ve_0 \big(r_t^j r_t^l\big)\right|.
\ee
The distance between the different rooms $R_t^j$
is greater or equal to $\rho_t$ according to
 (\ref{rom}).
Then, by Lemma \ref{l5.3'}, i)
 and condition {\bf V2},
\beqn\la{i222}
I_{22}&\le&
\sum\limits_{j< l}
| \langle Q_\ve(x,y),\chi_t^j\Phi_r(x,t)\otimes
\chi_t^l\Phi_r(y,t)\rangle|\nonumber\\
&\le&Ct^{-d}\sum\limits_{j< l}
\sum\limits_x \chi_t^j(x)
\sum\limits_y\chi_t^l(y) (1+|x-y|)^{-\gamma}
\nonumber\\
&\sim&
 t^{-d}n_t^2 t^{d-1}\Delta_t
\int\limits_{\rho_t}^{+\infty}(1+s)^{-\gamma}s^{d-1}\,ds
\sim n_t (1+\rho_t)^{-\gamma+d},
\eeqn
which vanishes as $t\to\infty$ because of
 (\ref{7.15'}) and $\gamma>d$.
Finally, it remains to check  that $I_{23}\to 0$,
$t\to\infty$. We have
$$
{\cal Q}_{\ve,t,r}(\Psi,\Psi)
=\mathbb{E}^\ve_0\big(\langle Y_0,\Phi_r(\cdot,t)\rangle^2\big)
=\mathbb{E}^\ve_0\big(\Big({\sum}_j (r_t^j+c_t^j)+l_t\Big)^2\big),
$$
according  to (\ref{razli}).
Therefore, by the Cauchy-Schwarz inequality,
\beqn
I_{23}&\le& \Big|
\mathbb{E}^\ve_0\big(\bigl({\sum}_j r_t^j\bigr)^2\big)
- \mathbb{E}^\ve_0\big(\bigl({\sum}_j r_t^j +
{\sum}_j c_t^j+l_t\bigr)^2 \big)\Big|\nonumber\\
& \le&
C n_t{\sum}_j \mathbb{E}^\ve_0 \big(|c_t^j|^2\big)  +
C_1\Bigl(
\mathbb{E}^\ve_0\big(({\sum}_j r_t^j)^2\big)\Bigr)^{1/2}
\Bigl(
n_t{\sum}_j \mathbb{E}^\ve_0\big(|c_t^j|^2\big)+
\mathbb{E}^\ve_0 \big(|l_t|^2\big)\Bigr)^{1/2}\nonumber\\
&&+C  \mathbb{E}^\ve_0 \big(|l_t|^2\big).\la{107}
\eeqn
Then  (\ref{106}), (\ref{i22}) and (\ref{i222}) imply
\beqn
\mathbb{E}^\ve_0\big(({\sum}_j r_t^j)^2\big)&\le&
{\sum}_j \mathbb{E}^\ve_0\big(|r_t^j|^2\big)
 \!+\!2{\sum}_{j<l}\Big|\mathbb{E}^\ve_0\big( r_t^j r_t^l\big)\Big|
\nonumber\\
&\le&
Cn_t\De_t/t+C_1n_t(1+\rho_t)^{-\gamma+d}\le C_2<\infty.
\nonumber
\eeqn
Now (\ref{106''}), (\ref{106'''}), (\ref{107}) and (\ref{7.15'}) yield
\beqn\nonumber
I_{23}\le C_1  n_t^2\rho_t/t+C_2 n_t(\rho_t/t)^{1/2}
+C_3 t^{-p} \to 0,~~t\to \infty.
\eeqn
So,  the terms $I_{21}$, $I_{22}$, $I_{23}$
in  (\ref{104}) tend to zero.
Then  (\ref{104}) implies that for $t=\tau/\ve$
\be\la{108}
I_2\le \frac{1}{2}\,
\left|{\sum}_j \mathbb{E}^\ve_0\big(|r_t^j|^2\big)-
 {\cal Q}_{\tau,r}(\Psi, \Psi)\right|
\to 0,~~\ve\to0.
\ee
{\it Step (iii)}
It remains to verify that for $t=\tau/\ve$
$$
I_3=\Big|\mathbb{E}^\ve_0\big(\exp\big\{i{\sum}_j r_t^j\big\}\big)
-\exp\big\{
-\fr12 {\sum}_j \mathbb{E}^\ve_0\big(|r_t^j|^2\big)\big\}\Big|
 \to 0,~~\ve\to0.
$$
Condition {\bf V3} yields
\beqn
&&\Big|\mathbb{E}^\ve_0\big(\exp\{i{\sum}_j r_t^j\}\big)
-\prod\limits_{-n_t}^{n_t-1}
\mathbb{E}^\ve_0\big(\exp\{i r_t^j\}\big)\Big|\nonumber\\
&\le&
\Big|\mathbb{E}^\ve_0\big(\exp\{ir_t^{-n_t}\}
\exp\{i\sum\limits_{-n_t+1}^{n_t-1} r_t^j\}\big)  -
 \mathbb{E}^\ve_0\big(\exp\{ir_t^{-n_t}\}\big)\mathbb{E}^\ve_0
\big(\exp\{i\sum\limits_{-n_t+1}^{n_t-1} r_t^j\}\big)\Big|
\nonumber\\
&&+
\Big|\mathbb{E}^\ve_0\big(\exp\{ir_t^{-n_t}\}\big)
\mathbb{E}^\ve_0\big(\exp\big\{i\sum\limits_{-n_t+1}^{n_t-1} r_t^j\big\}\big)
-\prod\limits_{-n_t}^{n_t-1}
\mathbb{E}^\ve_0\big(\exp\{i r_t^j\}\big)\Big|
\nonumber\\
&\le& C(1+\rho_t)^{-\kappa}+
\Big|\mathbb{E}^\ve_0\big(\exp\big\{i\sum\limits_{-n_t+1}^{n_t-1} r_t^j\big\}\big)
-\prod\limits_{-n_t+1}^{n_t-1}
\mathbb{E}^\ve_0\big(\exp\{i r_t^j\}\big)\Big|.\nonumber
\eeqn
We then apply condition {\bf V3} recursively
and obtain, according to Lemma \ref{r},
$$
\Big|\mathbb{E}^\ve_0\big(\exp\{i{\sum}_j r_t^j\}\big)-
\prod\limits_{-n_t}^{n_t-1}
\mathbb{E}^\ve_0\big(\exp\{i r_t^j\}\big)\Big|
\le
C n_t(1+\rho_t)^{-\kappa}\Big|_{t=\tau/\ve}\to 0,\quad \ve\to0.
$$
It remains to check that for $t=\tau/\ve$
$$
\Big|\prod\limits_{-n_t}^{n_t-1} 
\mathbb{E}^\ve_0\big(\exp\{ir_t^j\}\big)
-\exp\big\{-\fr12{\sum}_j \mathbb{E}^\ve_0\big(|r_t^j|^2\big)\big\}
\Big|
 \to 0,~~\ve\to0.
$$
According to the standard statement of the
Central Limit Theorem (see, e.g., \ci[Theorem 4.7]{P}),
it suffices to verify the  Lindeberg condition:
$\forall\de>0$,
$$
\left.\frac{1}{\sigma_t}
{\sum}_j \mathbb{E}_0^{\ve,\de\sqrt{\sigma_t}}
\big(|r_t^j|^2\big)\right|_{t=\tau/\ve} \to 0,~~\ve\to0.
$$
Here
$\sigma_t= {\sum}_j \mathbb{E}^\ve_0 \big(|r^j_t|^2\big)$,
and $\mathbb{E}_0^{\ve,a} (f)\equiv \mathbb{E}^\ve_0 (X^a f)$,
where $X^a$ is the indicator of the
event $|f|>a^2.$ Note that (\ref{108})
and (\re{5.*}) imply  that $\sigma_{\tau/\ve}\to
{\cal Q}_{\tau,r}(\Psi, \Psi)\not= 0,$
$\ve\to0.$
Hence it remains to verify that
\be\la{linc}
{\sum}_j \mathbb{E}_0^{\ve,a}
\big(|r_{\tau/\ve}^j|^2\big) \to 0,~~\ve\to0,
 ~~ \mbox{ for any }\, a>0.
\ee
We check Eqn (\ref{linc}) in Section 9.
This will complete the proof of Proposition \ref{l2.2}.
\hfill$\bo$

\setcounter{equation}{0}
\section{The Lindeberg condition}
The proof of (\ref{linc})
 is reduced  to the proof of
the following convergence
\be\la{111}
{\sum}_j \mathbb{E}^\ve_0\big(|r_{\tau/\ve}^{j}|^4 \big)\to 0,~~\ve\to0,
\ee
by using Chebyshev's inequality.
We deduce (\ref{111}) from
the following lemma.
\begin{lemma}  \la{p5.1}
Let the conditions  of Theorem \ref{the3}  hold.
Then for any $\Psi\in {\cal D}^0$
the following bounds hold,
\be\la{112}
\mathbb{E}^\ve_0\big(|r_t^{j}|^4\big)\le
C(\Psi) \De_t^2/t^2,   ~~t>1.
\ee
\end{lemma}
{\bf Proof.} {\it Step 1}~
Given four points $x^1,x^2,x^3,x^4\in\Z^d$, we set\\
$M_\ve^{(4)}(x^1,...,x^4)=
\mathbb{E}^\ve_0\left(Y(x^1)\otimes...\otimes Y(x^4)\right)$.
Then, similarly to (\ref{100rq})
we have
\be
\mathbb{E}^\ve_0\big(|r_t^{j}|^4\big)=
\langle \chi_t^j(x^1) \ldots \chi_t^j(x^4)M_\ve^{(4)}(x^1,\dots,x^4),
\Phi_r(x^1,t)\otimes\dots\otimes\Phi_r(x^4,t)\rangle.
\ee
Lemma \ref{l5.3'}, i) implies
\be\la{500}
\mathbb{E}^\ve_0\big(|r_t^{j}|^4\big)\le
 Ct^{-2d}\sum\limits_{i=2}^4
\sum\limits_{ \bar x\in(\Z^d)^4 }
\chi_t^j(x^1) \ldots \chi_t^j(x^4)
 |{\rm M}^{(4)}_\ve(\bar x)|.
\ee
By condition {\bf V4}, we have
\beqn\la{501}
\sum\limits_{ \bar x\in(\Z^d)^4 }
\chi_t^j(x^1) \ldots \chi_t^j(x^4)
 |{\rm M}^{(4)}_\ve(\bar x)|\le
\sum\limits_{(i_1,i_2,i_3,i_4)\in P\{1,2,3,4\}}
V_{i_1,i_2,i_3,i_4}(t),
\eeqn
where
\beqn\nonumber
V_{i_1,i_2,i_3,i_4}(t)= C \sum\limits_{\bar x}
\chi_t^j(x^1) \ldots \chi_t^j(x^4)
  (1+|x^{i_1}-x^{i_2}|)^{-\gamma}
(1+|x^{i_3}-x^{i_4}|)^{-\gamma}.
\eeqn
Similarly to  (\ref{100rq}), we have
\beqn
V_{i_1,i_2,i_3,i_4}(t)&\le& C
\sum\limits_{ x^{i_1}}\chi_t^j(x^{i_1})
\sum\limits_{x^{i_2} }
\chi_t^j(x^{i_2}) (1+|x^{i_1}-x^{i_2}|)^{-\gamma}
\nonumber\\
&&\times\sum\limits_{x^{i_3}}\chi_t^j(x^{i_3})
\sum\limits_{x^{i_4}} \chi_t^j(x^{i_4})
(1+|x^{i_3}-x^{i_4}|)^{-\gamma} \nonumber\\
&\sim& \Big[\De_t t^{d-1}
\sum\limits_{x^{i_2}} (1+|x^{i_1}-x^{i_2}|)^{-\gamma}\Big]^2. \nonumber
\eeqn
The  sum in $x^{i_2}$ is bounded
since $\gamma>d$. Hence,
\be\la{115}
 V_{i_1,i_2,i_3,i_4}(t)
\le C   \Delta^2_t t^{2d-2}.
\ee
Now the estimate (\ref{112}) follows from (\ref{500}),
(\ref{501}) and  (\ref{115}).
This completes the proof of Lemma \ref{p5.1}.
\hfill$\bo$


\end{document}